\documentclass[10 pt]{article}

\usepackage{media9}
\usepackage{comment}
\usepackage{amsmath } 
\usepackage{amssymb }
\usepackage{amsthm }  
\usepackage{mathtools}
\usepackage{graphicx} 
\usepackage{wrapfig} 
\usepackage{dsfont}
\usepackage{physics}
\usepackage{enumitem} 
\usepackage{bm}
\usepackage{hyperref}
\usepackage{xcolor}
\usepackage[margin = .5 in]{geometry}

\newcommand\noneq{\text{neq}}

\newcommand\bb[1]{\mathbb{#1}}
\newcommand\bbE{\mathbb{E}}
\newcommand\bbP{\mathbb{P}}
\newcommand\bbEb[1]{\bbE\left[ #1 \right]}

\newcommand\diff{\text{d}}
\newcommand\E{\mathcal{E}}
\newcommand\Laplace{\Delta}
\newcommand\fF{\mathcal{F}}
\newcommand\fL{\mathcal{L}}
\newcommand{\fH}{\mathcal{H}}
\newcommand\BigO{\mathcal{O}}
\newcommand\Indic{\mathds{1}}

\newcommand\avg[1]{\left\langle #1 \right\rangle}
\newcommand{\pd}{\partial}
\newcommand\mat[1]{\begin{pmatrix} #1 \end{pmatrix}}

\begin{document}

\begin{center}
    \LARGE
    A statistical mechanics framework for constructing non-equilibrium thermodynamic models \\ 
    \vspace{1 cm}
    \large
    Travis Leadbetter$^\text{a}$, Prashant K. Purohit$^\text{b}$, and Celia Reina$^\text{b,1}$\\
    September 2023\\
    \vspace{1 cm}
    \footnotesize
    {}$^\text{a}$Graduate Group in Applied Mathematics and Computational Science, University of Pennsylvania, Philadelphia, PA, 19104;\\
    {}$^\text{b}$Department of Mechanical Engineering and Applied Mechanics, University of Pennsylvania, Philadelphia, PA, 19104;\\{}$^\text{1}$To whom correspondence should be addressed: creina@seas.upenn.edu
\end{center}

\begin{abstract}
    Far-from-equilibrium phenomena are critical to all natural and engineered systems, and essential to biological processes responsible for life. For over a century and a half, since Carnot, Clausius, Maxwell, Boltzmann, and Gibbs, among many others, laid the foundation for our understanding of equilibrium processes, scientists and engineers have dreamed of an analogous treatment of non-equilibrium systems. But despite tremendous efforts, a universal theory of non-equilibrium behavior akin to equilibrium statistical mechanics and thermodynamics has evaded description. Several methodologies have proved their ability to accurately describe complex non-equilibrium systems at the macroscopic scale, but their accuracy and predictive capacity is predicated on either phenomenological kinetic equations fit to microscopic data, or on running concurrent simulations at the particle level. 
Instead, we provide a framework for deriving stand-alone macroscopic thermodynamics models directly from microscopic physics without fitting in overdamped Langevin systems. 
The only necessary ingredient is a functional form for a parameterized, approximate density of states, in analogy to the assumption of a uniform density of states in the equilibrium microcanonical ensemble. 
We highlight this framework's effectiveness by deriving analytical approximations for evolving mechanical and thermodynamic quantities in a model of coiled-coil proteins and double stranded DNA, thus producing, to the authors’ knowledge, the first derivation of the governing equations for a phase propagating system under general loading conditions without appeal to phenomenology.
The generality of our treatment allows for application to any system described by Langevin dynamics with arbitrary interaction energies and external driving, including colloidal macromolecules, hydrogels, and biopolymers. 
\end{abstract}

\section*{Significance}
The beautiful connection between statistical mechanics and equilibrium thermodynamics is one of the crowning achievements in modern physics. Significant efforts have extended this connection into the non-equilibrium regime. Impactful, and in some cases surprising, progress has been achieved at both the macroscopic and microscopic scales, but a key challenge of bridging these scales remains. In this work, we provide a framework for constructing macroscopic non-equilibrium thermodynamic models from microscopic physics without relying on phenomenology, fitting to data, or concurrent particle simulations. 
 We demonstrate this methodology on a model of coiled-coil proteins and double stranded DNA, producing the first analytical approximations to the governing equations for a phase transforming system without phenomenological assumptions.
\section*{Introduction}
nderstanding and predicting far-from-equilibrium behavior is of critical importance for advancing a wide range of research and technological areas including dynamic behavior of materials, \cite{jaeger2010far,complexDynamicsBAA}, complex energy systems \cite{hemminger2007directing}, as well as geological and living matter \cite{connolly2009geodynamic,gompper2020}. 
Although our understanding of each of these diverse fields continues to grow, a universal theory of non-equilibrium processes has remained elusive. 
The past century, however, has seen numerous significant breakthroughs towards this ultimate goal, of which we detail only a few below. 
At the macroscopic scale, classical irreversible thermodynamics leverages the local equilibrium assumption to allow classical thermodynamic quantities to vary over space and time, enabling one to describe well known linear transport equations such as Fourier's and Fick's laws \cite{lebon2008understanding}. Extended irreversible thermodynamics further promotes the fluxes of these quantities to the level of independent variables in order to capture more general transport laws \cite{jou1996}. Further extensions to allow for arbitrary state variables (not just fluxes), or history dependence take the names of thermodynamics with internal variables (TIV) or rational thermodynamics, respectively \cite{maugin1994,maugin1994thermodynamics2,coleman1964,truesdell1984historical}. More recently, the General Equation for Non-Equilibrium Reversible-Irreversible Coupling (GENERIC) framework and Onsager's variational formalism have proven to be successful enhancements of the more classical methods \cite{grmela1997dynamics,onsager1931reciprocal,doi2011onsager,mielke2011formulation}. 
On the other hand, linear response theory and fluctuation dissipation relations constitute the first steps towards a theory of statistical physics away from equilibrium. In the last few decades, interest in microscopic far-from-equilibrium processes has flourished due to the unforeseen discovery of the Jarzynski equality and other fluctuation theorems, as well as the advent of stochastic thermodynamics \cite{jarzynski1997,crooks1999,seifert2005,seifert2012,horowitz2020}, and the application of large deviation theory to statistical physics \cite{feng2006large,peletier2014variational,mielke2016generalization}. These advances have changed the way scientists view thermodynamics, entropy, and the second law particularly at small scales. 
\par 
 More specific to this work is the challenge of uniting scales. Given the success of the aforementioned macroscopic thermodynamic theories, how can one derive and inform the models within them  using microscopic physics? Describing this connection constitutes the key challenge in formulating a unified far-from-equilibrium theory. As of yet, the GENERIC framework possesses the strongest microscopic foundation. Starting from a Hamiltonian system, one can either coarse grain using the projection operator formalism \cite{ottinger1998} or a statistical lack-of-fit optimization method \cite{turkington2013,pavelka2020generalization} in order to derive the GENERIC equations. However, these methods are either challenging to implement, analytically or numerically, or contain fitting parameters which must be approximated from data. Alternatively, one can begin from a special class of stochastic Markov processes and use fluctuation-dissipation relations or large deviation theory to the same effect \cite{li2019harnessing,montefusco2021framework}. So far, numerical implementations of these methods have only been formulated for purely dissipative systems, with no reversible component.   
\par 
For this work, we shall utilize the less stringent framework of TIV, but recover GENERIC in an important case utilized in the examples. We will show how to leverage a variational method proposed by Eyink \cite{eyink1996} for evolving approximate non-equilibrium probability distributions to derive the governing equations of TIV for systems whose microscopic physics is well described by Langevin dynamics. Furthermore, in the approach proposed here, the variational parameters of the probability density are interpreted as macroscopic internal variables, with dynamical equations fully determined through the variational method. Once the approximate density is inserted into the stochastic thermodynamics framework, the equations for the classical macroscopic thermodynamics quantities including work rate, heat rate, and entropy production appear naturally, and possess the TIV structure. For example, the internal variables do not explicitly appear in the equation for the work rate, and the entropy production factors into a product of fluxes and their conjugate affinities, which themselves are given by the gradient of a non-equilibrium free energy. 
Moreover, we show that when the approximating density is assumed to be Gaussian, the internal variables obey a gradient flow dynamics with respect to the non-equilibrium free energy, and so the resulting rate of entropy production is guaranteed to be non-negative.
This direct link between microscopic physics and TIV has not been elaborated elsewhere, and we refer to this method as stochastic thermodynamics with internal variables (STIV).
\par
\begin{figure}[h!]
    \begin{center}
        \includegraphics{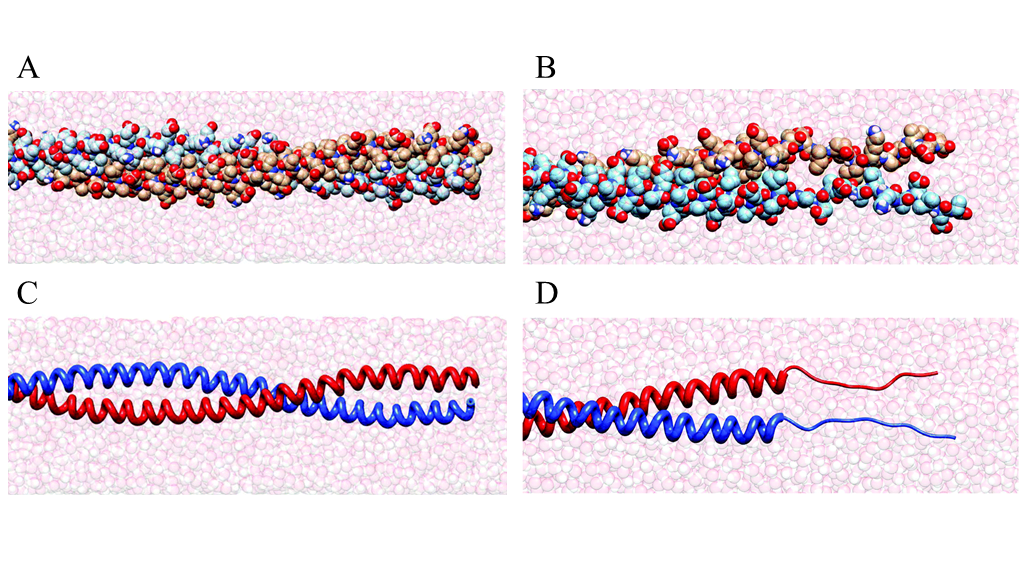}
        \caption{The stochastic thermodynamics with internal variables (STIV) framework proposed here provides kinetic and thermodynamic equations for a broad class of systems described by Langevin dynamics, including the coiled-coil protein depicted in these snapshots. Taken from molecular dynamics simulations, atomic level structures are depicted in (A) and (B), while the unfolding due to an externally applied load becomes clear in the secondary structures shown in (C) and (D). Vital for the coiled-coil protein's function, we study the dynamics of this transition from folded to unfolded configuration as a demonstration of the power of the STIV framework. Reproduced from \cite{torres2019combined} Fig.\ 1 with permission from the Royal Society of Chemistry.}
        \label{fig:coiledcoil}
    \end{center}
\end{figure}

To illustrate and highlight the effectiveness of this method, we provide the results of two examples. The first is a paradigmatic example from stochastic thermodynamics: a single colloidal particle acted on by a linear external force, mimicking a macromolecule in an optical trap. 
It demonstrates all of the key features of the method while being simple enough to allow for comparison to exact solutions.
The second example features a model system for studying phase transitions of bio-molecules, for example in coiled-coil proteins \cite{kreplak2004new,torres2019combined} (depicted in Fig.\ \ref{fig:coiledcoil}) or double stranded DNA \cite{gore2006,van2009unraveling}: a colloidal mass-spring-chain system with double-well interactions between neighboring masses. By comparing to Langevin simulations, we show that STIV not only produces accurate analytical approximations to relevant thermodynamic quantities, but also predicts the speed of a traveling phase front induced by external driving.

\section*{Theory} 
\subsection*{Stochastic thermodynamics} 
We begin by outlining the key ideas of stochastic thermodynamics which defines classical thermodynamic quantities at the trajectory level for systems obeying Langevin dynamics, such as those embedded in an aqueous solution. 
These quantities include work, heat flow, and entropy production among others, and these new definitions allow for an expanded study of far-from-equilibrium behavior at the level of individual, fluctuating trajectories. Stochastic thermodynamics is a highly active area of study, and has been developed far beyond what is detailed here, as we have limited our presentation to only what we need for introducing STIV. See \cite{seifert2012} and the references therein for further details.
\par
The paradigmatic example within stochastic thermodynamics is a colloidal particle in a viscous fluid at constant temperature, $T$, acted on by an external driving (we present the theory for a single particle in one dimension as the generalization to many particles in multiple dimensions is straightforward). 
This system is well described by an overdamped Langevin equation, which can be written as a stochastic differential equation of the form
\begin{equation*}
     \text{d}x(t) = -\frac1{\eta}\frac{\pd e}{\pd x}(x,\lambda)\, \text{d}t + \sqrt{2d}\, \text{d}b(t),
\end{equation*}
where $x(t)$ denotes the particle's position at time $t \in [t_\text{i},t_\text{f}]$, $\eta$ is the drag coefficient of the particle in the fluid, $- \frac{\pd e}{\pd x}(x,\lambda)$ is the force acting on the particle coming from a potential energy, $e$, $\lambda(t)$ is a prescribed external control protocol, $d = \frac1{\eta\beta}$ is the diffusion coefficient, $\beta = 1/k_BT$ the inverse absolute temperature in energy units, and $b(t)$ is a standard Brownian motion. 
\par
Given this system, stochastic thermodynamics enables one to define the internal energy, work, heat, and entropy at the level of the trajectory. Naturally, $e(x(t),\lambda(t))$ defines the internal energy of the system. One does work on the system by changing $e$ via the external control, $\lambda$. Thus, the incremental work reads
\begin{equation} \label{eqn:work_incr}
 \diff w = \frac{\pd  e}{\pd \lambda}\ \dot{\lambda}\,\diff t.
\end{equation}
Using the first law of thermodynamics, we conclude that the incremental heat flowing out of the system is 
\begin{equation*}
\diff q = \diff w - \diff e. 
\end{equation*}

\par
An additional important quantity is the total entropy, $s^{\text{tot}}$. 
From the second law of thermodynamics, its macroscopic counterpart, $S^\text{tot}$ (to be defined), should be non-decreasing and describe the level of irreversiblity of the trajectory. 
To that end, the change in total entropy is defined using the log of the (Raydon-Nikodym) derivative of the probability of observing the given trajectory, $\bbP[x(t)\mid \lambda]$, with respect to the probability of observing the reversed trajectory under the time reversed external protocol, $\tilde{\bbP}[\tilde{x}(t)\mid \tilde{\lambda}]$
\begin{equation*}
\Delta s^{\text{tot}}[x(t)] = k_B \log(\frac{\diff \bbP[x(t)\mid \lambda]}{\diff\tilde{\bbP}[\tilde{x}(t)\mid \tilde{\lambda}]}) 
\end{equation*}
where $\tilde{x}(t) = x(t_\text{f} - t)$ and likewise for $\tilde{\lambda}$. Upon taking the expectation with respect to all possible trajectories (and any probabilistic initial conditions), 
\begin{equation*}
    \Delta S^{\text{tot}} = \avg{\Delta s^\text{tot}}_{\text{paths}} = \int \Delta s^\text{tot}[x(t)]\,\diff \bbP[x(t)\mid \lambda]
\end{equation*} 
is recognized as $k_B$ times the Kullback-Leibler divergence between the distributions of forward and backwards trajectories. As such, $\Delta S^{\text{tot}}$ must be non-negative.
It is also useful to break up the total entropy change into the change in the entropy of the system, 
\begin{equation*}
    \Delta s[x(t)] = -k_B\log(\frac{p(x(t_\text{f}),t_\text{f}\mid \lambda)}{p(x(t_\text{i}),t_\text{i}\mid\lambda)}),
\end{equation*} 
where $p(x,t\mid \lambda)$ is the probability density of observing the particle at position $x$ at time $t$, and the change in the entropy of the medium
\begin{equation} \label{eqn:def_sm} 
\Delta s^{\text{m}} = \Delta s^\text{tot} - \Delta s.
\end{equation}
\par
Finally, one defines the microscopic non-equilibrium free energy in terms of the potential and entropy as $a^\noneq = e - Ts$ \cite{still2012thermodynamics}. 
Using the path integral representation of $\bbP[x(t)\mid \lambda]$ and $\tilde{\bbP}[\tilde{x}(t)\mid \tilde{\lambda}]$, one finds that the incremental heat dissipated into the medium equals the incremental entropy change in the medium $T \diff s^\text{m} =  \diff q$ \cite{seifert2008stochastic}. This allows one to relate the change in non-equilibrium free energy to the work done and the change in total entropy
\begin{align}
    \diff a^{\noneq} &= \diff e - T\diff s\nonumber\\
    &= \diff w - \diff q - T \diff s\nonumber\\
    \label{eqn:def_stot}
    &= \diff w - T\diff s^{\text{tot}}.
\end{align}
\par
As we saw with $\Delta S^\text{tot}$, each microscopic quantity has a macroscopic counterpart defined by taking the expectation with respect to all possible paths. Throughout, we use the convention that macroscopic (averaged) quantities are written in capital, and microscopic quantities are written in lower case, e.g., $A^{\noneq} = \avg{a^{\noneq}}_{\text{paths}}$.

\subsection*{Thermodynamics with internal variables} 
Now we turn to the macroscopic description, and give a brief overview of Thermodynamics with internal variables (TIV). TIV has enjoyed decades of application as an important tool of study for irreversible processes in solids, fluids, granular media, and viscoelastic materials \cite{ortiz1999variational,nemat2004plasticity,simo2006computational,gurtin2010mechanics,dunatunga2015continuum}. 
Originally formulated as an extension to the theory of irreversible processes, TIV posits that non-equilibrium description without history dependence requires further state variables beyond the classical temperature, number of particles, and applied strain (in the canonical ensemble, for example) in order to determine the system's evolution \cite{maugin1994,horstemeyer2010}. 
These additional variables, the internal variables, encode the effects of the microscopic degrees of freedom on the observable macrostate. 
Thus, the relevant state functions take both classical and internal variables as input. 
The flexibility of the theory is apparent from the wide range of material behavior it can describe. 
The challenge, however, is in selecting descriptive internal variables, and in defining their kinetic equations in a way which is consistent with microscopic physics. Here, we take on the latter challenge. 
\subsection*{Variational method of Eyink} 
The key mathematical tool we utilize for connecting TIV to stochastic thermodynamics is a variational method for approximating non-equilibrium systems laid out by Eyink \cite{eyink1996}. 
This method generalizes the Rayleigh-Ritz variational method of quantum mechanics to non-Hermitian operators. 
The method assumes the system in question can be described by a probability density function governed by an equation of the form $\frac{\pd}{\pd t} p = \fL p$ (e.g., a Fokker-Planck equation associated with Langevin particle dynamics). 
Since the operator $\fL$ is not Hermitian, $\fL \neq \fL^\dagger$, one must define a variational method over both probability densities $p$ and test functions $\psi$. 
Begin by defining the non-equilibrium action functional 
\begin{equation*}
    \Gamma[\psi,p] = \int_0^\infty \int_X \psi( \frac{\pd}{\pd t} - \fL )p\,\diff x\, \diff t.
\end{equation*}
Under the constraint that 
\begin{equation*}
    \int_X \psi\ p\,\diff x \Big|_{t = \infty} = \int_X \psi\ p\,\diff x\Big|_{t = 0},
\end{equation*}
this action is stationary, $\delta \Gamma[\psi^*,p^*] = 0$, if and only if $(\frac{\pd}{\pd t} - \fL)p^* = 0$ and $(\frac{\pd}{\pd t} + \fL^\dagger)\psi^* = 0$. 
By defining the non-equilibrium ``Hamiltonian'' $\fH[\psi,p] = \int_X \psi\ \fL p\ dx$, one can recast the variational equation $\delta \Gamma[\psi^*,p^*] = 0$ in Hamiltonian form
\begin{align}
    \label{eqn:variation_p}
    \frac{\pd}{\pd t} p^* &= \frac{\delta}{\delta \psi} \fH[\psi^*,p^*] \\
    \label{eqn:variation_psi}
    \frac{\pd}{\pd t} \psi^* &= - \frac{\delta}{\delta p} \fH[\psi^*,p^*].
\end{align}
As it stands, the variation is taken over two infinite dimensional function spaces, and as such, it is only possible to find exact solutions in a handful of systems. 
However, one can still make use of these dynamical equations to find a variational approximation to the true solution which lies within some fixed subspace. 
To do so, one begins by assuming the true density, $p^*(x,t)$, and test function $\psi^*(x,t)$, can be approximated by a parameterized density $\hat{p}(x,\alpha(t))$ and test function $\hat{\psi}(x,\alpha(t))$ respectively, so that all of the time dependence is captured by the variables $\alpha(t) = (\alpha_1(t),...,\alpha_N(t))$. 
For example, a standard method for choosing a parameterization is to pick an exponential family \cite{casella2021statistical}, or specifically a collection of quasi-equilibrium distributions \cite{turkington2013}. In this case, one selects a finite number of linearly independent functions of the state $\{\phi_i(x)\}_{i=1}^N$ to serve as observables describing the system. The parameterized densities $\hat{p}(x,\alpha(t))$ are defined as (for time dependent ``natural'' parameters $\alpha(t)$) 
\begin{equation*}
\hat{p}(x,\alpha(t)) = \exp( \sum_{i=1}^N \alpha_i(t)\phi_i(x) + \fF(\alpha(t)))
\end{equation*}
where $\fF(\alpha) = -\log(\int \exp(\sum_{i=1}^N\alpha_i \phi_i(x))\,\diff x)$ is a log-normalizing constant. The primary reason for using this parameterization is that for each $\alpha$, this $\hat{p}(x,\alpha)$ has maximum Shannon entropy with respect to all other probability densities subject to the constraint that the averages $\avg{\phi_i(x)}_{\hat{p}}$ take on prescribed values. In the quasi-equilibrium case, $\phi_1(x)$ is almost always taken as the system energy, and hence $\alpha_1(t)$ becomes $\beta$.  
\par 
Given any parameterization, quasi-equilibrium or otherwise, the dynamical equations Eq.\ \ref{eqn:variation_p} and Eq.\ \ref{eqn:variation_psi} reduce to a coupled system of ordinary differential equations (ode)
\begin{equation}
    \label{eqn:variational_ode}
    \big\{\alpha_i,\alpha_j\big\}\frac{\diff \alpha_j}{\diff t} = \frac{\pd \fH}{\pd \alpha_i}
\end{equation}
where 
\begin{equation*}
    \big\{\alpha_i,\alpha_j\big\} = \int_X \frac{\pd \hat{\psi}}{\pd \alpha_i} \frac{\pd \hat{p}}{\pd \alpha_j} - \frac{\pd \hat{\psi}}{\pd \alpha_j}\frac{\pd \hat{p}}{\pd \alpha_i} \,\diff x.
\end{equation*}
The solution to Eq.\ \ref{eqn:variational_ode}, $\alpha^*(t)$, offers the best approximations to the true solution $p^*(x,t) \approx \hat{p}(x,\alpha^*(t))$, $\psi^*(x,t) \approx \hat{\psi}(x,\alpha^*(t))$, lying within the parameterized subspace. 
\subsection*{Stochastic thermodynamics with internal variables} 
Finally, we fuse stochastic thermodynamics with this variational framework to provide a general method for constructing TIV models. 
Stochastic thermodynamics provides the appropriate thermodynamic definitions, while the variational formalism of Eyink will allow us to derive dynamical equations for the internal variables consistent with the microscopic physics. 
\par
We return to the colloidal particle system with governing stochastic differential equation
\begin{equation*}
    \diff x(t) = -\frac1{\eta}\frac{\pd e}{\pd x} (x,\lambda)\ \diff t + \sqrt{2d}\ \diff b(t).
\end{equation*}
If $p(x,t\mid \lambda)$ is the probability density of observing the system in state $x$ at time $t$ given a prespecified external protocol, $\lambda(t)$, then $p(x,t\mid \lambda)$ obeys the Fokker-Planck equation
\begin{equation*}
    \frac{\pd p}{\pd t}  = \fL\ p = \frac1{\eta} \frac{\pd}{\pd x} \cdot (\frac{\pd e}{\pd x}\ p) + d \Laplace_x p. 
\end{equation*}
When $\lambda(t)$ is held constant, the true density tends towards the equilibrium Boltzmann distribution, $p^*(x,t\mid \lambda) \propto \exp(-\beta e(x,\lambda))$. Away from equilibrium, $p^*(x,t\mid \lambda)$ may be highly complex, and in that case we would like to find a low dimensional representation which captures the physical phenomena of interest. To do so, we choose a class of parameterized densities $\hat{p}(x,\alpha)$ to use in the variational method of Eyink, keeping in mind that the variables $\alpha(t)$ are to become the internal variables in the macroscopic description. This is in direct analogy with the assumption of a uniform probability density in the microcanonical ensemble, or the Maxwellian distribution in the canonical ensemble. Note, also that in keeping with ensembles in which volume or strain is controlled rather than force or stress, we assume no explicit dependence on the external protocol $\lambda$ in $\hat{p}(x, \alpha)$. This will prove necessary mathematically in what follows. Finally, we do not explicitly consider the dependence of $\hat{p}$ on $\beta$, as we have assumed that temperature is constant.  
\par
We next define the approximate entropy $\hat{s}(x,\alpha) = -k_B\log(\hat{p}(x,\alpha))$ and use its derivatives with respect to the internal variables to define the test functions in the variational formalism
\begin{equation*}
    \hat{\psi}(x,\alpha,\gamma) = 1 + \gamma \cdot \frac{\pd \hat{s}}{\pd \alpha}.
\end{equation*}
Since the true solution to the adjoint equation $\frac{\pd \psi^*}{\pd t} = -\fL^\dagger \psi^*$ is $\psi^* \equiv \text{const.}$, the variables $\gamma$ serve as expansion coefficients about the true solution $\psi^* \equiv 1$. In the \nameref{SI}, we show that they essentially function as dummy variables, as the variational solution fixes $\gamma(t) \equiv 0$ for all time.
Hence, the vector $\alpha(t)$ will be the only relevant variable. 
Assuming this choice of density and test functions, the variational formalism of Eyink yields the dynamical equation 
\begin{equation}
    \label{eqn:alpha_dyn}
    \avg{\frac{\pd \hat{s}}{\pd \alpha} \frac{\pd \hat{s}}{\pd \alpha}^T}_{\hat{p}}\cdot \dot{\alpha} = -k_B\avg{\fL^\dagger \frac{\pd \hat{s}}{\pd \alpha}}_{\hat{p}}
\end{equation}
where $\avg{g}_{\hat{p}} = \int g(x)\hat{p}(x,\alpha)dx$ denotes averaging with respect to $\hat{p}$. This equation reveals the utility of our choice of $\hat{\psi}$. The matrix on the left hand side $\mathbb{F}_{ij} = \avg{ \frac{\pd \hat{s}}{\pd \alpha_i} \frac{\pd \hat{s}}{\pd \alpha_j}}_{\hat{p}}$ is $k_B^2$ times the Fisher information matrix of the density $\hat{p}(x,\alpha)$ \cite{turkington2013}. This matrix is always symmetric, and is positive definite so long as the functions $\{\frac{\pd \hat{s}}{\pd \alpha_i}(x,\alpha)\}_{i=1}^N$ are linearly independent as functions of $x$ for all $\alpha$. Picking $\alpha(0)$ such that $\hat{p}(x,\alpha(0)) \approx p^*(x,0\mid \lambda)$, and using Eq.\ \ref{eqn:alpha_dyn} to solve for $\alpha(t)$ gives us the variational solution for $\hat{p}(x,\alpha(t)) \approx p^*(x,t\mid \lambda)$ for all time. 
\par 
Having approximated the density using the internal variables, we turn to stochastic thermodynamics to impose the thermodynamic structure. In order to make use of the approximate density, $\hat{p}$, we simply use the stochastic thermodynamics definitions of thermodynamic quantities at the macroscale, but make the substitution $p^*(x,t\mid \lambda) \rightarrow \hat{p}(x,\alpha(t))$. Following this rule, we generate the thermodynamic quantities as 
\begin{align}  
    \hat{E}(\alpha,\lambda) &= \avg{e}_{\hat{p}} \nonumber \\
    \hat{S}(\alpha) &= -k_B\avg{\log(\hat{p})}_{\hat{p}} \nonumber\\
    \hat{A}^{\noneq}(\alpha,\lambda) &= \hat{E} - T\hat{S} \nonumber\\
    \label{eqn:aprx_workrate}
    \frac{d}{dt}\hat{W}(\alpha,\lambda) &= \avg{\frac{\pd e}{\pd \lambda}\dot{\lambda}}_{\hat{p}} \\
    \label{eqn:aprx_Stot}
    T\frac{d}{dt}\hat{S}^\text{tot}(\alpha,\lambda) &= \frac{d}{dt}\hat{W} - \frac{d}{dt}\hat{A}^{\noneq} \\
    \label{eqn:aprx_Sm}
    \frac{d}{dt}\hat{S}^\text{m}(\alpha,\lambda) &= \frac{d}{dt}\hat{S}^\text{tot} - \frac{d}{dt}\hat{S}
\end{align}
where Eq.\ \ref{eqn:aprx_workrate}, \ref{eqn:aprx_Stot}, and \ref{eqn:aprx_Sm} are derived from Eq.\ \ref{eqn:work_incr}, \ref{eqn:def_stot}, and \ref{eqn:def_sm} respectively, as shown in the \nameref{SI}. 
Since we have assumed a constant bath temperature for the governing Langevin equation, we do not explicitly write the dependence of the quantities above on $\beta$.
Recall, a key assumption is that the approximate density should be independent of $\lambda$ for fixed $\alpha$. Hence, the approximate entropy, $\hat{S}$, is a function of $\alpha$ alone. This means that the partial derivative with respect to $\lambda$ can be factored out of the expectation in Eq.\ \ref{eqn:aprx_workrate}. Since $\hat{S}$ does not depend on $\lambda$, we may write 
\begin{equation*}
    \frac{d}{dt} \hat{W} = \frac{\pd \hat{E}}{\pd \lambda} \dot{\lambda} = \frac{\pd}{\pd \lambda} \left(\hat{E} - T\hat{S} \right)\dot{\lambda} = \frac{\pd \hat{A}^{\noneq}}{\pd \lambda}  \dot{\lambda} \equiv \hat{F}^\text{ex}\dot{\lambda},
\end{equation*}
so that the approximate external force is given by the gradient of $\hat{A}^\noneq$ with respect to the external protocol, $\hat{F}^\text{ex} \equiv \frac{\pd \hat{A}^{\noneq}}{\pd \lambda}$. 
Moreover, Eq.\ \ref{eqn:aprx_Stot} and Eq.\ \ref{eqn:aprx_Sm} simplify to 
\begin{equation*}
    T\frac{d}{dt}\hat{S}^\text{tot} = -\frac{\pd \hat{A}^{\noneq}}{\pd \alpha}\dot{\alpha} \qquad\qquad  
    \frac{d}{dt} \hat{Q} = - \frac{\pd \hat{E}}{\pd \alpha}\dot{\alpha}.
\end{equation*}
Thus, the approximate work rate and the approximate rate of entropy production of the medium are given by the derivatives of $\hat{E}$ and the approximate work rate and the approximate rate of total entropy production are given by the derivatives of $\hat{A}^{\noneq}$.
In particular, the rate of total entropy production takes the form of a product of fluxes, $\dot{\alpha}$, and affinities, $\mathcal{A}_\alpha = -\frac{\pd \hat{A}^{\noneq}}{\pd \alpha}$. Likewise, the internal variables do not explicitly enter into the equation for the work rate, just as in TIV. 
Moreover, in the \nameref{SI}, we prove that for an arbitrary interaction energy $e(x,\lambda)$, internal variables obey the stronger GENERIC structure \cite{ottinger2005}, obeying a gradient flow equation with respect to the non-equilibrium free energy, whenever the approximate probability density is assumed to be Gaussian. In this case, the internal variables are the mean and inverse covariance ($\alpha = (\mu,\Sigma^{-1})$) of the probability density of the state, $x \in \bb{R}^N$.
Symbolically, we define
\begin{equation}\label{eqn:phat_gauss}
    \hat{p}(x, \mu,\Sigma^{-1}) = \sqrt{\det(\frac{\Sigma^{-1}}{2\pi})}\exp(-\frac1{2}(x-\mu)^T\Sigma^{-1}(x-\mu)).
\end{equation}
This choice of form for the approximate density is a standard choice in popular approximation methods including Gaussian phase packets \cite{heller1975time,gupta2021nonequilibrium} and diffusive molecular dynamics \cite{kulkarni2008variational,li2011diffusive} primarily for its tractable nature.

As mentioned, the dynamics of $\mu$ and $\Sigma^{-1}$ are given in terms of gradients with respect to the non-equilibrium free energy
\begin{equation} \label{eqn:grad_flow}
    \dot{\mu} = -\frac1{\eta}\frac{\pd \hat{A}^{\noneq}}{\pd \mu}, \qquad\dot{\Sigma}^{-1} = - M(\Sigma^{-1}) : \frac{\pd \hat{A}^{\noneq}}{\pd \Sigma^{-1}}
\end{equation}
for a positive semi-definite dissipation tensor $M(\Sigma^{-1})$, and hence, the total rate of entropy production is guaranteed to be non-negative
\begin{equation}\label{eqn:ent_prod}
    T\frac{d}{dt}\hat{S}^{\text{tot}} = \frac1{\eta} \norm{\frac{\pd \hat{A}^\noneq}{\pd \mu}}^2 + \frac{\pd \hat{A}^\noneq}{\pd \Sigma^{-1}} : M : \frac{\pd \hat{A}^\noneq}{\pd \Sigma^{-1}}.
\end{equation}
Thus, we see that the thermodynamic structure emerges naturally by utilizing the variational method of Eyink within the context of stochastic thermodynamics, and that we are not forced to postulate phenomenological equations for $\alpha(t)$. They emerge directly from the variational structure. 

\section*{Results} 
\subsection*{A single colloidal particle} 
To illustrate the STIV framework we apply it to a toy model: an overdamped, colloidal particle acted on by an external force that is linear in the extension of a spring connected to the particle. Despite its simplicity, this model is often used to describe a molecule caught in an optical trap. 
In one dimension, the governing Langevin equation for the particle's position is given by $\text{d}x = -\frac1{\eta} \frac{\pd e}{\pd x}(x,\lambda)\text{d}t + \sqrt{2 d}\, \text{d}b$, where $e(x,\lambda) = \frac{k}{2}(\lambda- x)^2$ is the energy of the spring or the trapping potential, and $\lambda(t)$ is an arbitrary external protocol. The corresponding Fokker-Planck operator is $\fL\ p = \frac1{\eta}\frac{\pd}{\pd x}\left( \frac{\pd e}{\pd x} p \right) + d \frac{\pd^2}{\pd x^2} p$. The true solution is an Ornstein-Uhlenbeck (O.U.) process, thus, providing an exactly solvable model for comparison \cite{steele2001}. Since the probability density of the O.U. process is Gaussian for all time (assuming a Gaussian initial distribution), we use a Gaussian approximate distribution with mean $\mu$ and standard deviation $\sigma$ as internal variables (Eq.\ \ref{eqn:phat_gauss} with $\Sigma^{-1} = 1/\sigma^2$). 
It is straightforward to input this density into the variational formalism of Eyink and compute the dynamics. 
The details of the derivation are written out in the \nameref{SI}. The resulting dynamical equations recover those of the O.U. process
\begin{equation*}
    \dot{\mu} = -\frac{k}{\eta}(\mu - \lambda),\qquad
    \dot{\sigma} = -\frac{k}{\eta}\sigma\left(1 - \frac1{k\beta\sigma^2}\right).
\end{equation*}
\par
Now that we have the dynamics, we turn to computing the thermodynamics quantities. Of particular interest is the fact that the fluxes of the internal variables are linear in the affinities, $-\frac{\pd \hat{A}^{\noneq}}{\pd \mu} = \eta\dot{\mu}$, $-\frac{\pd \hat{A}^{\noneq}}{\pd \sigma} = \eta\dot{\sigma}$, 
hence ensuring a non-negative entropy production. 
We can also find the approximate work rate, heat rate, and rate of total entropy production explicitly
\begin{equation*}
    \frac{\diff}{\diff t}\hat{W} = \eta \dot{\mu}\dot{\lambda}, \quad\quad \frac{\diff}{\diff t}\hat{Q} = \eta \dot{\mu}^2 - k\sigma\dot{\sigma}, \quad\quad T\frac{\diff}{\diff t}\hat{S}^{\text{tot}} = \eta \dot{\mu}^2 + \eta \dot{\sigma}^2. 
\end{equation*}
\begin{figure}[h!]
    \begin{center}
        \includegraphics{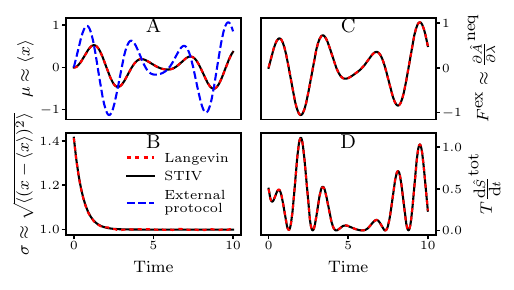}
        \caption{A comparison of the STIV method (black solid line) to Langevin simulations (red dashes, 100,000 simulations) for a single colloidal particle in a harmonic optical trap. (A) The mean mass position, $\mu \approx \avg{x}$, as well as the external pulling protocol, $\lambda(t)$, in blue. (B) The standard deviation, $\sigma \approx \sqrt{\avg{(x-\avg{x})^2}}$, of mass positions. (C) The external force on the optical trap. (D) The total rate of entropy production.}
        \label{fig:single}
    \end{center}
\end{figure}
\par
Although a toy system, this example highlights the fact that when the true solution to the governing PDE for the probability density lies in the subspace spanned by the trial density, the true solution is recovered and relevant thermodynamic quantities can be exactly computed via the non-equilibrium free energy, as can be seen in Fig.\ \ref{fig:single}. 
\subsection*{Double-well colloidal mass-spring-chain} 
\begin{figure*}[ht!]
    \begin{center}
        \includegraphics[width = 17.7 cm]{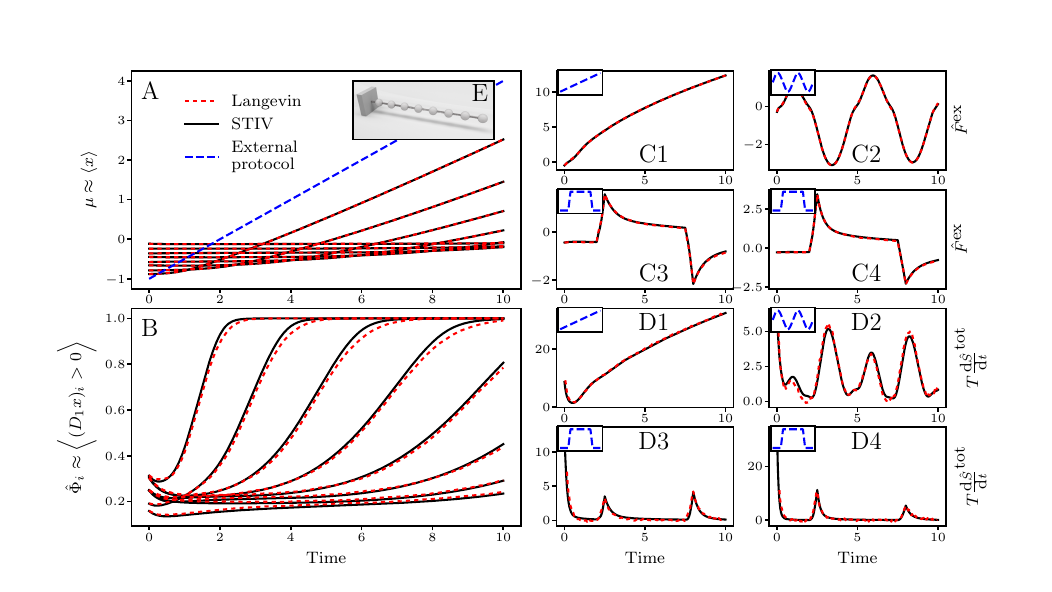}
        \caption{(A) A comparison of the predicted mean mass locations using STIV (black lines) and empirical mean of 100,000 Langevin simulations (red dashes) for the 8 mass colloidal mass-spring-chain with double-well interactions and a linear external protocol (external protocol shown in blue throughout). Except in (C4) and (D4), the parameters of the symmetric interaction potential are $k_1 = k_2 = l_1 = l_2 = 1$. (B) The predicted and simulated phase fractions of springs in the right well for the same system as (A). (C) The predicted versus simulated external force for four different pulling protocols: (1) linear, (2) sinusoidal, (3) step, (4) step with an asymmetric interaction potential between masses ($k_1 = 1, l_1 = 1, k_2 = 2, l_2 = 1/2$). (D) The predicted versus simulated rate of total entropy production for the same four pulling protocols as in (C). The external protocols used are shown in the insets of (C,D). (E) Cartoon of the mass-spring-chain configuration. One side is held fixed while the other is controlled by the external protocol.}
        \label{fig:dw}
    \end{center}
\end{figure*}

For our primary example, we study a colloidal mass-spring-chain system with double-well interaction between masses. 
Depicted in the inset of Fig.\ \ref{fig:dw} E, this model of phase front propagation in coiled-coil proteins and double stranded DNA contains several metastable configurations corresponding to the different springs occupying one of the two minima in the interaction energy,
and exhibits phase transitions between them.
A key test for the STIV framework is whether or not the phase can accurately be predicted, and more importantly, whether the kinetics and thermodynamics of phase transitions can be captured without phenomenological kinetic equations. An almost identical model to the one studied here is considered in \cite{truskinovsky2005kinetics}, but in a Hamiltonian setting rather than as a colloidal system. Here, the authors make use of the piece-wise linearity of the force, $-\frac{\pd e}{\pd x}$, to derive an exact solution for the strain in the presence of a phase front traveling at constant velocity, and the kinetic relation for this phase front without the use of phenomenological assumptions. Our solution, on the other hand, is inherently approximate (though accurate), but does not depend on either the assumptions of constant velocity of the phase front, or the specific piece-wise linear form of the force. The choice of interaction potential is simply convenience, and the STIV method could be easily applied to quartic or other double-well interaction potentials.
\par
We assume each spring has internal energy described by the following double well potential:
\begin{equation*}
    \label{eqn:dwPot}
    u(z) = \begin{cases} \frac{k_1}{2}(z + l_1)^2& x \leq 0 \\ \frac{k_2}{2}(z - l_2)^2 + h_2  & x > 0 \\ \end{cases} 
\end{equation*}
where $h_2$ is chosen so that $u(z)$ is continuous (i.e., $h_2 = (k_1 l_1^2 - k_2 l_2^2)/2$). 
For simplicity, we have placed one well on each side of the origin so that the transition point falls at $z = 0$. 
Letting $x = (x_1,...,x_N)$ be the positions of the $N$ interior masses, the total energy, given an external protocol $\lambda$, is $e(x,\lambda) = \sum_{i=1}^N u(x_i - x_{i-1}) + u(\lambda - x_N)$ where $x_0 \equiv 0$.  
\par
We begin by assuming that the positions of the masses can be well described using a multivariate Gaussian distribution, and set the internal variables to be the mean $\mu$ and the inverse covariance $\Sigma^{-1}$ as in Eq.\ \ref{eqn:phat_gauss}. The exact form of the dynamical equations for the internal variables induced by the STIV framework can be found in the \nameref{SI}. As expected, the equations obey the gradient flow structure given by Eq.\ \ref{eqn:grad_flow}, where in this case we have $M_{ij,kl} = \frac1{\eta}(\Sigma^{-1}_{ik}\Sigma^{-2}_{jl} + \Sigma^{-2}_{ik}\Sigma^{-1}_{jl} + \Sigma^{-1}_{il}\Sigma^{-2}_{jk} + \Sigma^{-2}_{il}\Sigma^{-1}_{jk})$. 
The rate of total entropy production, given by Eq.\ \ref{eqn:ent_prod}, is thus non-negative. It is interesting to note that the dynamical equations for $\mu$ and $\Sigma^{-1}$ are coupled through an approximation of the phase fraction of springs occupying the right well
\begin{equation*}
    \hat{\Phi}_i(x,t) \equiv \int_{-\infty}^\infty \Indic_{(x_i - x_{i-1} > 0)}\, \hat{p}(x,\mu(t),\Sigma^{-1}(t)) \diff x.
\end{equation*} 
As an important special case, fixing the interaction parameters to produce a quadratic interaction, $l_1 = - l_2$ and $k_1 = k_2 = k$, causes the dependence on $\hat{\Phi}$ to drop out, and the equations from $\mu$ and $\Sigma^{-1}$ decouple.
\begin{figure*}[h!]
    \centering
    \includegraphics{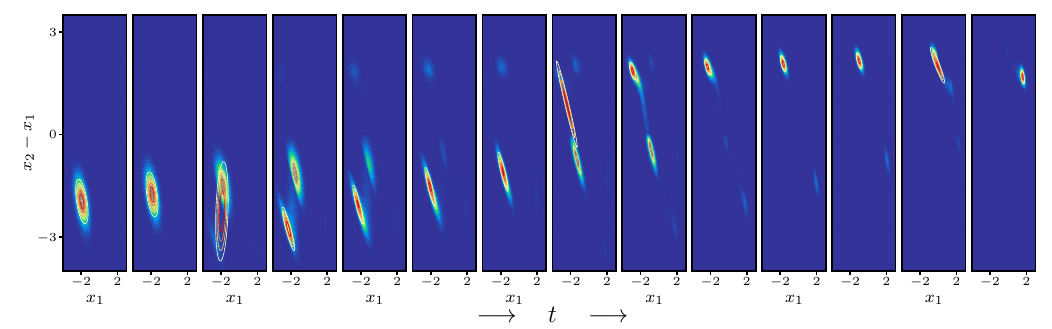}
    \caption{A comparison of the probability density for the spring lengths for a two mass mass-spring-chain system with double-well spring energies.  The colored histograms depict densities collected from 100,000 Langevin simulations of the solution to the governing stochastic differential equation, while the grey-scale contour lines show the approximation using STIV. On each panel, the horizontal axis gives the length of the first spring, $x_1$, and the vertical axis gives the length of second, $x_2 - x_1$. Panels from left to right show equal increments in time. We see that despite missing the details of the multi-modal behavior apparent in the Langevin simulations, the STIV approximation successfully tracks the location and size of the dominant region of non-zero probability.}
    \label{fig:density}
\end{figure*}
\par
In Fig.\ \ref{fig:density}, we show a comparison of the probability densities produced by the STIV framework for a two mass system to those obtained from Langevin simulations of the governing stochastic differential equation. Although fine details of the multimodal structure are missed (as is to be expected when using a Gaussian model), the size and location of the dominant region of non-zero probability is captured, making it possible to compute the relevant macroscopic thermodynamic quantities, as we discuss next.  
\par
Since the exact form of the true solution $p^*(x,t\mid \lambda)$ is unknown, we compare the results of the framework to simulations of the Langevin dynamics of a system with 8 free masses in Fig.\ \ref{fig:dw}.
Despite the fact that the true solution is multimodal due to the existence of several metastable configurations, it's clear that the approximations of the mean mass position (A), phase fraction (B), external force ($\frac{\pd E}{\pd \lambda} \approx \frac{\pd \hat{E}}{\pd \lambda} = \frac{\pd \hat{A}^\noneq}{\pd \lambda}$) (C), and total rate of entropy production (D) are all highly accurate. This holds true for a variety of pulling protocols including linear (1), sinusoidal (2), and a step displacement (3,4), as well as for symmetric (1,2,3) and asymmetric (4) interaction potentials. Returning to (B), we see that for a system with an initial configuration in which all the springs begin in the left well we can observe a propagating phase front as the springs, one by one, transition from the left to the right well. This transition is captured by the internal variable model with high accuracy allowing one to directly approximate the velocity of the phase front. We note, however, that the quantitative accuracy of the method appears to hold most strongly in the case that the thermal energy is significantly larger or smaller than the scale of the energy barrier separating the two potential energy wells in the spring interaction. When the thermal energy and potential energy barriers are at the same scale, the true density of states is highly multimodal, and not well approximated by a multivariate Gaussian, see \nameref{Movie S2}. In this case, the STIV approximation captures the behavior of only the dominant mode. When the thermal energy is large relative to the barrier, the thermal vibrations cause the modes to collapse into a single ``basin'' which can be well approximated by the STIV density, see \nameref{Movie S1}. Finally, when the thermal energy is small, the true density is unimodal, and undergoes rapid jumps between the different energy minima. The Gaussian STIV density, again, becomes an effective choice for approximation.   
\par
The dynamical equations for the internal variables take the form of a discretized partial differential equation (pde). Assuming we properly rescale the parameters of the interaction potential, the viscosity, and temperature so that the equilibrium system length, energy, entropy, and quasistatic viscous dissipation are independent of the number of masses ($l_i = l_i^0/N$, $k_i= Nk_i^0$, $\eta = \eta^0/N$, $\beta = N\beta^0$ ($i \in \{1,2\}$)) then, in the limit as the number of masses tends to infinity the internal variables $\mu_i$ and $\Sigma^{-1}_{ij}$ become functions of continuous variables $x \in [0,1]$ and $x,y \in [0,1]\times[0,1]$, respectively. Since it is challenging to invert a continuum function $\Sigma^{-1}(x,y,t)$, we make use of the identity $\dot{\Sigma}_{ij} = -(\Sigma \dot{\Sigma}^{-1} \Sigma)_{ij}$ to derive the following limiting pde for $\mu(x,t)$,  $\Sigma(x,y,t)$, the strain, $\epsilon(x,t) \equiv \frac{\pd \mu}{\pd x}(x,t)$, and the covariance of the strain, $\E(x,y,t) \equiv \frac{\pd^2 \Sigma}{\pd x\pd y}(x,y,t)$ 
\begin{align*}
    \frac{\pd \mu}{\pd t} &= \frac1{\eta_0}\frac{\pd}{\pd x} \bigg\{ k_1^0\left( \epsilon + l_1^0\right)(1 - \hat{\Phi}) + k_2^0\left(\epsilon - l_2^0\right)\hat{\Phi} + (k_2^0 - k_1^0)\E \frac{\pd \hat{\Phi}}{\pd \epsilon} \bigg\}\\
    \frac{\pd \Sigma}{\pd t} &= 2\Laplace^w \Sigma  \\
    \\
    &\mu(x = 0,t) = 0,  \qquad \mu(x = l_0,t) = \lambda(t) \\
    &\Sigma(x=0,y,t) = \Sigma(x=l_0,y,t) = 0 \\
    &\Sigma(x,y=0,t)=\Sigma(x,y=l_0,t) = 0\\
\end{align*}
with the approximate phase fraction defined through 
\begin{equation*}
    \hat{\Phi}(x,t) = \hat{\Phi}(\epsilon,\E) = \Phi\left(\frac{\epsilon(x,t)}{\sqrt{\E(x,x,t)}}\right).
\end{equation*}
Here, $\Laplace^w = \frac{\pd }{\pd x}w(x,t) \frac{\pd }{\pd x} + \frac{\pd }{\pd y}w(y,t) \frac{\pd }{\pd y}$, ${w(x,t) = \frac{k_1^0}{\eta^0}(1- \hat{\Phi}) + \frac{k_2^0}{\eta^0}\hat{\Phi} - \frac1{\eta^0}(k_1^0l_1^0 + k_2^0l_2^0)\frac{\pd \hat{\Phi}}{\pd \epsilon}}$, and $\Phi(\xi)$ is the cumulative distribution function of a standard Gaussian (mean zero, variance one). Both equations for $\frac{\pd \mu}{\pd t}$ and $\frac{\pd \Sigma}{\pd t}$ contain contributions from the left well (the terms multiplying $(1- \hat{\Phi})$), the right well (the terms multiplying $\hat{\Phi}$), and the phase boundary (the terms multiplying $\frac{\pd \hat{\Phi}}{\pd \epsilon}$), and in the \nameref{SI} we give assumptions on the continuum limit for $\Sigma(x,y,t)$ such that these dynamical equation maintain the gradient flow structure
\begin{align*}
    \frac{\pd \mu}{\pd t} &= -\frac1{\eta}\frac{\delta \hat{A}^\noneq}{\delta \mu} \\ 
    \frac{\pd \sigma}{\pd t} &= - \int_0^1\int_0^1 M(x,y,z,w,t)\frac{\delta \hat{A}^\noneq}{\delta \Sigma}(z,w,t) \diff z\diff w.
\end{align*}
\par
\begin{figure}[h!]
    \begin{center}
        \includegraphics{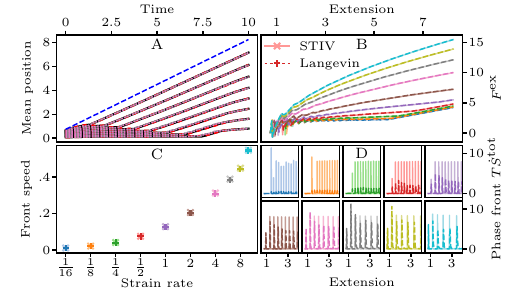}
        \caption{(A) Mean mass positions for Langevin and STIV approximations to an 17 mass (Langevin: red dashes, STIV: solid black) and a 62 mass (Langevin: pink short dashes, STIV: grey long dashes) double-well mass-spring-chain system, with parameters rescaled for the same effective behavior. For both systems, only the 8 masses expected to overlap are plotted. Throughout (B,C,D), darker colors, dashed lines, and + scatter points denote results from Langevin simulations, whereas lighter colors, solid lines, and x scatter points denote results from the STIV approximation. (B) The external force as a function of extension for the 17 mass system at ten different strain rates (shown in (C)). (C) The phase front speed as a function of strain rate in the 17 mass system. (D) The rate of entropy production due to the phase front as a function of extension for each of the strain rates shown in (C).}
        \label{fig:continuum}
    \end{center}
\end{figure}

In Fig.\ \ref{fig:continuum} (A), we demonstrate that the continuum response of the system can be well approximated through the STIV framework with finitely many masses. We see agreement between the mean mass positions observed in Langevin simulations and those predicted using the STIV framework for both 17 and 62 masses, verifying that both discretizations capture the continuum response. This allows us to use the 17 mass system to accurately predict important continuum level quantities such as the external force as a function of extension, $\lambda$, \ref{fig:continuum} (B), the phase front speed, \ref{fig:continuum} (C), for different applied strain rates, and finally the rate of entropy production due to the phase front, \ref{fig:continuum} (D), as a function of the system extension for each of the strain rates shown in (C).
Methods for computing the front speed and the rate of entropy production due to the phase front can be found in the \nameref{SI}.
\par 
Finally, in the continuum limit, one can differentiate in time the defining equation for the location of the phase front in the reference configuration, $\hat{\Phi}(\hat{I}(t),t) \equiv \frac1{2}$ to yield the following ordinary differential equation for the location of the phase front
\begin{equation*}
    \frac{\diff}{\diff t}\hat{I}(t) = -\frac{\frac{\pd^2}{\pd x^2} \frac{\delta \hat{A}^\noneq}{\delta \epsilon}(x,t) }{\eta \frac{\pd^2 \mu}{\pd x^2}(x,t)} \Big|_{x = \hat{I}(t)}.
\end{equation*} 
This equation reveals that the phase front is directly proportional to the ratio of the curvature of the thermodynamic affinity conjugate to the strain $\mathcal{A}_\epsilon \equiv - \frac{\delta \hat{A}^{\noneq}}{\delta \epsilon}$ and the curvature of $\mu$ at the location of the phase front.

\section*{Discussion} 
Our results demonstrate the utility and accuracy of the STIV framework as a method for constructing TIV models which are consistent with microscopic physics. 
After assuming a functional form for a set of parameterized probability densities which serve to approximate the true density of states, inserting this approximation into the thermodynamic definitions taken from stochastic thermodynamics directly yields the internal variables structure, and the dynamics of these internal variables are fully determined by the variational method of Eyink. 
The resulting macroscopic model encodes the microscopic features of the system to the degree allowed within the provided probability density without any need for further reference back to smaller scales.
Moreover, in the important case of a Gaussian form for the approximate probability density, $\hat{p}(x,\alpha)$, we recover the gradient flow dynamics and the GENERIC structure which is commonly assumed without direct microscopic justification. 
In this work, we have focused on examples yielding analytically tractable approximations. 
However, it is equally possible to extend the method beyond such constraints by creating a numerical implementation based on sampling techniques using modern statistical and machine learning techniques.
Furthermore, extensions to Hamiltonian systems, active noise and models exhibiting significant coarse graining constitute important future steps for the STIV framework.

\section*{Acknowledgment}
T.L.~acknowledges that this project was supported in part by a fellowship award under contract FA9550-21-F-0003 through the National Defense Science and Engineering Grauate (NDSEG) Fellowship Program, sponsored by the Air Force Research Laboratory (AFRL), the Office of Naval Research (ONR) and the Army Research Office (ARO). P.K.P.~acknowledges support from ACS, USA grant number PRF-61793 ND10. C.R.~gratefully acknowledges support from NSF CAREER Award, CMMI-2047506.


\section*{SI Appendix}\label{SI}
\subsection{Derivation of dynamics for STIV}
Here, we derive the dynamical equations for the internal variables using the variational method of Eyink \cite{eyink1996}. We use the functional form $\hat{p} = \hat{p}(x,\alpha)$ and $\hat{\psi}(x,\alpha,\gamma) = 1 + \gamma \cdot \frac{\pd \hat{s}}{\pd \alpha}$ as described in the main text, where $\hat{s}(x,\alpha) = -k_B\log(\hat{p}(x,\alpha))$.
\par
Recall that the dynamical equations for the vector of variables $(\alpha,\gamma)$ is given by 
\begin{align}
    \{\alpha_i,\alpha_j\}\dot{\alpha}_j + \{\alpha_i,\gamma_j\}\dot{\gamma}_j &= \frac{\pd \fH}{\pd \alpha_i} \label{eqn:EyinkDyn1} \\
     \{\gamma_i,\alpha_j\}\dot{\alpha}_j + \{\gamma_i,\gamma_j\}\dot{\gamma}_j &= \frac{\pd \fH}{\pd \gamma_i}
    \label{eqn:EyinkDyn2}
\end{align}
where the parameterized non-equilibrium Hamiltonian is given by $\fH(\alpha,\gamma) = \int_X \hat{\psi} \fL \hat{p}\, \diff x$ and the bracket by 
\begin{equation*}
    \{\mu_i,\nu_j\} = \int \frac{\pd \hat{\psi}}{\pd \mu_i}\frac{\pd \hat{p}}{\pd \nu_j} - \frac{\pd \hat{\psi}}{\pd \nu_j}\frac{\pd \hat{p}}{\pd \mu_i} \,\diff x 
\end{equation*}
for $\mu,\nu = \alpha$ or $\gamma$ (we shall use the Einstein summation convention throughout unless stated otherwise). Using the specific forms of $\hat{p}$ and $\hat{\psi}$ in the non-equilibrium Hamiltonian gives 
\begin{equation*}
\fH(\gamma,\alpha) = \int \left(1 + \gamma\cdot \frac{\pd \hat{s}}{\pd \alpha}(x,\alpha)\right) \fL \hat{p}(x,\alpha) \,\diff x.
\end{equation*}
Integrating by parts to transform $\fL$ into it's adjoint $\fL^\dagger$, noting that $\fL^\dagger 1 = 0$ and $\fL^\dagger$ is linear gives
\begin{align*}
 \fH(\gamma,\alpha) &= \int  \hat{p}\fL^\dagger \left(1 + \gamma\cdot \frac{\pd \hat{s}}{\pd \alpha}(x,\alpha)\right) \,\diff x \nonumber\\
    &= \int \hat{p} \fL^\dagger \frac{\pd \hat{s}}{\pd \alpha} \,\diff x \cdot \gamma \nonumber\\
    &= \avg{\fL^\dagger \frac{\pd \hat{s}}{\pd \alpha}}_{\hat{p}} \cdot \gamma. 
\end{align*}

\par 
Next we compute the bracket, $\{\gamma_i,\alpha_j\}$, 
\begin{align*}
    \{\gamma_i,\alpha_j\} &= \int \frac{\pd}{\pd \gamma_i} \left(1 + \gamma \cdot \frac{\pd \hat{s}}{\pd \alpha}\right) \frac{\pd \hat{p}}{\pd \alpha_j} - \frac{\pd}{\pd \alpha_j} \left(1 + \gamma \cdot \frac{\pd \hat{s}}{\pd \alpha}\right) \frac{\pd \hat{p}}{\pd \gamma_i} \,\diff x \nonumber\\
    &= \int \frac{\pd \hat{s}}{\pd \alpha_i}\frac{\pd \hat{p}}{\pd \alpha_j} \,\diff x \nonumber\\
    &= \int \frac{\pd \hat{s}}{\pd \alpha_i}\frac{\pd \log(\hat{p})}{\pd \alpha_j} \hat{p} \,\diff x \nonumber\\
    &= -\frac1{k_B} \avg{\frac{\pd \hat{s}}{\pd \alpha_i}\frac{\pd \hat{s}}{\pd \alpha_j}}_{\hat{p}}
\end{align*}
where between the first and second line we have used $\frac{\pd \hat{p}}{\pd \gamma} = 0$. 
For this same reason, we know $\{\gamma_i,\gamma_j\} = 0$ and $\frac{\pd \fH}{\pd \gamma} = \avg{\fL^\dagger \frac{\pd \hat{s}}{\pd \alpha}}_{\hat{p}}$. 
Using these formula in Eqn. \ref{eqn:EyinkDyn2} leaves us with
\begin{equation}
    \avg{\frac{\pd \hat{s}}{\pd \alpha_i}\frac{\pd \hat{s}}{\pd \alpha_j}}_{\hat{p}}\dot{\alpha}_j = -k_B \avg{\fL^\dagger \frac{\pd \hat{s}}{\pd \alpha}
_i}_{\hat{p}}.
    \label{eqn:alphaDyn}
\end{equation}
Since these equations do not depend on $\gamma$, they completely determine the dynamics of $\alpha$. However, for completeness, we derive the dynamical equations for $\gamma$ and show that $\gamma \equiv 0$ is a solution. 
\par 
We first note that $\{\alpha_i,\gamma_j\} = -\{\gamma_j,\alpha_i\}$ by definition. Next, we compute $\{\alpha_i,\alpha_j\}$ as 
\begin{align*}
    \{\alpha_i,\alpha_j\} &= \int_X \frac{\pd}{\pd \alpha_i} \left(1 + \gamma \cdot \frac{\pd \hat{s}}{\pd \alpha}\right) \frac{\pd \hat{p}}{\pd \alpha_j} - \frac{\pd}{\pd \alpha_j} \left(1 + \gamma \cdot \frac{\pd \hat{s}}{\pd \alpha}\right) \frac{\pd \hat{p}}{\pd \alpha_i} \,\diff x \nonumber\\
    &= -\frac1{k_B}\int_X \left(\gamma \cdot \frac{\pd}{\pd \alpha} \frac{\pd \hat{s}}{\pd \alpha_i}\right)\frac{\pd \hat{s}}{\pd \alpha_j}\hat{p} - \left(\gamma \cdot \frac{\pd}{\pd \alpha}\frac{\pd \hat{s}}{\pd \alpha_j}\right)\frac{\pd \hat{s}}{\pd \alpha_i}\hat{p} \,\diff x \nonumber \\ 
    &= -\frac1{k_B}\gamma_k \avg{ \left( \frac{\pd^2 \hat{s}}{\pd \alpha_i\pd \alpha_k}\frac{\pd \hat{s}}{\pd \alpha_j} - \frac{\pd \hat{s}}{\pd \alpha_i}\frac{\pd^2 \hat{s}}{\pd \alpha_j\pd\alpha_k}\right)}_{\hat{p}} 
\end{align*}
Finally,
\begin{equation*}
\frac{\pd \fH}{\pd \alpha_i} = \gamma_k \frac{\pd}{\pd \alpha_i} \avg{\fL^\dagger \frac{\pd \hat{s}}{\pd \alpha_k}}_{\hat{p}}.
\end{equation*}
Putting these all into Eqn. \ref{eqn:EyinkDyn1} gives
\begin{equation*}
    -\frac1{k_B}\gamma_k \avg{ \left( \frac{\pd^2 \hat{s}}{\pd \alpha_i\pd \alpha_k}\frac{\pd \hat{s}}{\pd \alpha_j} - \frac{\pd \hat{s}}{\pd \alpha_i}\frac{\pd^2 \hat{s}}{\pd \alpha_j\pd\alpha_k}\right)}_{\hat{p}}\dot{\alpha}_j + \frac1{k_B}\avg{\frac{\pd \hat{s}}{\pd \alpha_i} \frac{\pd \hat{s}}{\pd \alpha_j} }_{\hat{p}} \dot{\gamma}_j = \gamma_k \frac{\pd }{\pd \alpha_i}\avg{\fL^\dagger \frac{\pd \hat{s}}{\pd \alpha_k}}_{\hat{p}}.
\end{equation*}
After rearranging terms, we see that the equations have the form 
\begin{equation*}
M_{ij}(\alpha)\dot{\gamma}_j = Q_{ik}(\alpha,\dot{\alpha})\gamma_k
\end{equation*}
from which it is clear that $\gamma \equiv 0$ is a solution. 

\subsection{Approximate thermodynamic values in STIV}
We now justify the approximate equations 
\begin{align}  
    \frac{d}{dt}\hat{W}(\alpha,\lambda) &= \avg{\frac{\pd e}{\pd \lambda}\dot{\lambda}}_{\hat{p}} \label{eqn:aprx_work} \\
    T\frac{d}{dt}\hat{S}^\text{tot}(\alpha,\lambda) &= \frac{d}{dt}\hat{W} - \frac{d}{dt}\hat{A}^{\noneq} \\
    \frac{d}{dt}\hat{S}^\text{m}(\alpha,\lambda) &= \frac{d}{dt}\hat{S}^\text{tot} - \frac{d}{dt}\hat{S}
\end{align}
from equations
\begin{align}
    \diff w &= \frac{\pd e}{\pd \lambda}\dot{\lambda}\diff t \label{eqn:def_work}\\
    T\diff s^\text{tot} &= \diff a^{\noneq} - \diff w 
    \Delta s^\text{m} &= \Delta s^\text{tot} - \Delta s. 
\end{align}
The basic steps are to show that these relations hold for the averages, and then approximate the true density of states $p(x,t\mid \lambda)$ with $\hat{p}(x, \alpha)$. Starting with Eq.\ \ref{eqn:def_work}, we write out the stochastic differentials in integral form
\begin{equation*}
    w(t') - w(t) = \int_t^{t'} \frac{\pd e}{\pd \lambda}(x(s),\lambda(s))\dot{\lambda}\diff s.
\end{equation*}
We then average over all paths, divide by $\Delta t = t'-t$ and take the limit as $t' \rightarrow t^+$ to get 
\begin{align*}
    \lim_{t'\rightarrow t} \frac{\avg{w(t') - w(t)}_\text{paths}}{t'-t} &= \lim_{t'\rightarrow t} \frac1{t'-t}\avg{\int_t^{t'}\frac{\pd e}{\pd \lambda}(x(s),\lambda(s))\dot{\lambda}\diff s}_\text{paths} \\
    &=\lim_{t'\rightarrow t} \frac1{t'-t} \int_t^{t'}\avg{\frac{\pd e}{\pd \lambda}(x(s),\lambda(s))\dot{\lambda}}_\text{paths}\diff s \\
    &= \avg{\frac{\pd e}{\pd \lambda}(x(t),\lambda(t))\dot{\lambda}}_\text{paths}\\
    &= \avg{\frac{\pd e}{\pd \lambda}(x,\lambda(t))\dot{\lambda}}_{p}.
\end{align*}
Since $\lim_{t'\rightarrow t} \frac{\avg{w(t') - w(t)}_\text{paths}}{t'-t} = \frac{\diff}{\diff t}W(t)$, we recover Eq.\ \ref{eqn:aprx_work} by approximating $p$ with $\hat{p}$ on the right hand side
\begin{equation*}
    \frac{\diff}{\diff t}\hat{W}(t) \equiv \avg{\frac{\pd e}{\pd \lambda}(x,\lambda(t))\dot{\lambda}}_{\hat{p}}.
\end{equation*}
\par 
Moving on to Eq.\ \ref{eqn:def_stot}, we note the left hand side is again a finite difference, and we can simply take an expectation and use the definition of the derivative to get 
\begin{equation*}
    \lim_{t'\rightarrow t} \frac{T  \avg{\diff s^\text{tot}\big|_t^{t'}}_\text{paths}}{t'- t} = T\frac{\diff}{\diff t}S^\text{tot}. 
\end{equation*}
Taking the difference quotient on the right hand side yields
\begin{equation*}
    \lim_{t'\rightarrow t} \frac{\avg{ \diff a^{\noneq} - \diff w \big|_t^{t'} }_\text{paths}}{t'- t} = \frac{\diff}{\diff t}A^\noneq - \frac{\diff}{\diff t}W. 
\end{equation*}
We note that 
\begin{equation*}
    A^\noneq(t) = \avg{e(x(t),\lambda(t)) - \frac1{\beta}\log(p(x(t),t\mid \lambda))}_\text{paths} = \avg{e(x,\lambda(t)) - \frac1{\beta}\log(p(x,t\mid \lambda(t)))}_p,
\end{equation*}
and hence use
\begin{equation*}
    \hat{A}^\noneq(t) \equiv  \avg{e(x,\lambda(t)) - \frac1{\beta}\log(\hat{p}(x, \alpha(t) ))}_{\hat{p}(x,\alpha(t)}
\end{equation*}
to define
\begin{equation*}
    T\frac{\diff}{\diff t} \hat{S}^\text{tot} \equiv \frac{\diff}{\diff t}\hat{A}^\noneq - \frac{\diff}{\diff t} \hat{W}.
\end{equation*}
\par 
Finally, for Eq.\ \ref{eqn:def_sm} we can use 
\begin{equation*}
    S(t) = -k_B\avg{\log(p(x,t\mid \lambda(t)))}_p
\end{equation*}
to define 
\begin{equation*}
    \hat{S} \equiv -k_B\avg{\log(p(x\mid \alpha(t)))}_{\hat{p}(x, \alpha(t))}
\end{equation*}
and use this and the previous definition for $\frac{\diff}{\diff t}\hat{S}^\text{tot}$ to define $\frac{\diff}{\diff t}\hat{S}^\text{sm}$ analogously. 

\subsection{Gaussian approximation produces a gradient flow structure and non-negative entropy production for STIV} \label{sec:gradFlow}
Here we show that for an arbitrary interaction energy, the multivariate Gaussian approximation 
\begin{equation*}
    \hat{p}(x, \mu,\Sigma^{-1}) = \exp(-\frac1{2}(x - \mu)^T\Sigma^{-1}(x- \mu) + \frac1{2}\log(\det(\frac1{2\pi}\Sigma^{-1})))
\end{equation*}
to the solution of a colloidal system
\begin{equation*}
    \diff x(t) = -\frac1{\eta}\frac{\pd e}{\pd x}(x(t),\lambda(t))\ \diff t + \sqrt{2d}\ \diff b(t)
\end{equation*}
with constant drag $\eta$, inverse absolute temperature $\beta$, and diffusion coefficient $d = \frac1{\beta\eta}$ but arbitrary interaction energy, $e(x,\lambda)$,
produces a gradient flow dynamics for the internal variables, $\alpha =  (\mu,\Sigma^{-1})$, and as a result the approximate rate of total entropy production is guaranteed to be non-negative, $\frac{\diff}{\diff t}\hat{S}^\text{tot} \geq 0$ (since we are using a Gaussian approximation, we assume from the outset that $\Sigma^{-1}$ is symmetric; the resulting dynamical equations will maintain this symmetry for all time). 
\par 
We first compute the dynamical equations by computing Eqn. \ref{eqn:alphaDyn} for this system. We compute the derivatives of $\hat{s}$ with respect to the internal variables
\begin{align*}
    \frac{\pd \hat{s}}{\pd \mu_i} &= -k_B\Sigma^{-1}_{ij}(x - \mu)_j\\
    \frac{\pd \hat{s}}{\pd \Sigma^{-1}_{ij}} &= k_B \left(\frac{(x-\mu)_i(x-\mu)_j}{2} - \frac{\Sigma_{ij}}{2}\right). 
\end{align*}
From here, it is a straightforward computation to see 
\begin{align*}
    \avg{\frac{\pd \hat{s}}{\pd \mu_i} \frac{\pd \hat{s}}{\pd \mu_j}}_{\hat{p}} &= k_B^2\Sigma_{ij}^{-1} \\
    \avg{\frac{\pd \hat{s}}{\pd \mu_i}\frac{\pd \hat{s}}{\pd \Sigma^{-1}_{jk}}}_{\hat{p}} &= 0 \\
    \avg{\frac{\pd \hat{s}}{\pd \Sigma^{-1}_{ij}}\frac{\pd \hat{s}}{\pd \Sigma^{-1}_{kl}}}_{\hat{p}} & = \frac{k_B^2}{4}\left(\Sigma_{ik}\Sigma_{jl} + \Sigma_{il}\Sigma_{jk}\right) \equiv k_B^2 \ \mathbb{F}^{\Sigma^{-1}}_{ijkl}.\\
\end{align*}
The adjoint equation to the Fokker-Planck operator associated with the dynamics is $\fL^\dagger g = \frac1{\eta} f \cdot \frac{\pd g}{\pd x} + d \Laplace_x g$, where $f = -\frac{\pd e}{\pd x}$, so we must now compute 
\begin{align*}
    \frac{\pd^2 \hat{s}}{\pd x_i\pd \mu_j} &= -k_B \Sigma^{-1}_{ij}\\
    \frac{\pd^3 \hat{s}}{\pd x_i\pd x_i\pd \mu_j} &= 0\\
    \frac{\pd^2 \hat{s}}{\pd x_i\pd \Sigma^{-1}_{jk}} &= \frac{k_B}{2}\left(\delta_{ij}(x-\mu)_k + \delta_{ik}(x-\mu)_j\right)\\
    \frac{\pd^3 \hat{s}}{\pd x_i\pd x_i\pd \Sigma^{-1}_{jk}} &= k_B\delta_{jk}.\\
\end{align*}
After entering these equations into Eqn. \ref{eqn:alphaDyn} and simplifying the resulting expression, one gets
\begin{align}
    \dot{\mu} &=\frac1{\eta}\avg{f}_{\hat{p}} \label{eqn:MuDyn}\\
     \mathbb{F}^{\Sigma^{-1}}_{ijkl}\dot{\Sigma}^{-1}_{kl} &= -\left(\frac1{\eta}\text{sym}\left(\avg{f_i(x-\mu)_j}_{\hat{p}}\right) + d\ \delta_{ij}\right) \label{eqn:SigmaInvDyn}
\end{align}
Where $\text{sym}(M) = \frac1{2}(M + M^T)$ is the symmetric part of any matrix $M$. 
\par 
Next, we wish to start from the non-equilibrium free energy $\hat{A}^{\noneq}$ and show that it's derivatives can be related to these dynamical equations. We start from its definition 
\begin{equation*}
    \hat{A}^{\noneq} = \avg{e}_{\hat{p}} - T\avg{\hat{s}}_{\hat{p}} = \avg{e}_{\hat{p}} + \frac1{\beta}\avg{\log(\hat{p})}_{\hat{p}}.
\end{equation*}
For a Gaussian distribution 
\begin{equation*}
    \avg{\log(\hat{p})}_{\hat{p}} = -\frac{N}{2} + \frac1{2}\log(\det(\frac1{2\pi}\Sigma^{-1})),
\end{equation*}
where $N$ is the dimension of $x$ (i.e., $x \in \mathbb{R}^N$).
Taking derivatives of $\hat{A}^\noneq$ with respect to the internal variables gives
\begin{align}
    \frac{\pd \hat{A}^\noneq}{\pd \mu} &= \frac{\pd \avg{e}_{\hat{p}}}{\pd \mu} \label{eqn:dAdMu}\\
    \frac{\pd \hat{A}^\noneq}{\pd \Sigma^{-1}} &= \frac{\pd \avg{e}_{\hat{p}}}{\pd \Sigma^{-1}} + \frac1{2\beta}\Sigma. \label{eqn:dAdSigma}
\end{align}
Since $e$ doesn't explicitly depend on the internal variables, the derivatives only hit the probability density, 
\begin{equation*}
    \frac{\pd \avg{e}_{\hat{p}}}{\pd \alpha} = \int e \frac{\pd \hat{p}}{\pd \alpha} \diff x. 
\end{equation*}
Since $\hat{p}$ is Gaussian, derivatives of $\hat{p}$ with respect to $\mu$ and $\Sigma^{-1}$ can be equated to derivatives of $\hat{p}$ with respect to $x$. 
Clearly,
\begin{equation}
    \frac{\pd \hat{p}}{\pd \mu} = -\frac{\pd \hat{p}}{\pd x}.
    \label{eqn:derMu}
\end{equation}
Moreover,
\begin{equation*}
    \frac{\pd \hat{p}}{\pd \Sigma_{ij}^{-1}} = -\frac1{2}\left((x-\mu)_i(x-\mu)_j - \Sigma_{ij}\right)\hat{p}
\end{equation*}
and 
\begin{equation*}
    \frac{\pd^2 \hat{p}}{\pd x_i\pd x_j} = \frac{\pd}{\pd x_i}\left(-\Sigma^{-1}_{jk}(x-\mu)_k\hat{p}\right) = \left(\Sigma^{-1}_{il}\Sigma^{-1}_{jk}(x-\mu)_k(x-\mu)_l - \Sigma^{-1}_{ij}\right)\hat{p}.
\end{equation*}
We can then relate these two equations via
\begin{equation}
    \frac{\pd \hat{p}}{\pd \Sigma^{-1}_{ij}} = -\frac1{2}\Sigma_{ik}\Sigma_{jl} \frac{\pd^2\hat{p}}{\pd x_k\pd x_l} \label{eqn:derSigma}.
\end{equation}
\par
Now we use these relations to compute $\frac{\pd \hat{A}^\noneq}{\pd \mu}$ and $\frac{\pd \hat{A}^\noneq}{\pd \Sigma^{-1}}$.
First, for $\mu$, using \ref{eqn:dAdMu}, \ref{eqn:derMu}, integration by parts, and then \ref{eqn:MuDyn} shows
\begin{equation*}
    \frac{\pd \hat{A}^\noneq}{\pd \mu} = \int e \frac{\pd \hat{p}}{\pd \mu}\ \diff x = -\int e \frac{\pd \hat{p}}{\pd x}\ \diff x = \int \frac{\pd e}{\pd x} \hat{p}\ \diff x = -\avg{f}_{\hat{p}} = -\eta \dot{\mu}.
\end{equation*}
Thus, $\mu$ obeys a gradient flow equation with respect to $\hat{A}^\noneq$. 
On the other hand, for $\Sigma^{-1}$, following similar steps, we have 
\begin{align*}
    \frac{\pd \avg{e}_{\hat{p}}}{\pd \Sigma^{-1}_{ij}} &= \int e \frac{\pd \hat{p}}{\pd \Sigma^{-1}_{ij}} \ \diff x \\ 
    &= -\frac1{2}\Sigma_{ik}\Sigma_{jl} \int e \frac{\pd^2 \hat{p}}{\pd x_k\pd x_l} \ \diff x \\
    &=\frac1{2}\Sigma_{ik}\Sigma_{jl}\int \frac{\pd e}{\pd x_k}\frac{\pd \hat{p}}{\pd x_l} \ \diff x \\
    &=\frac1{2}\Sigma_{ik}\int f_k(x-\mu)_j\hat{p}\ \diff x \\ 
    &=\frac1{2}\Sigma_{ik} \avg{f_k(x-\mu)_j}_{\hat{p}}.
\end{align*}
Thus, we have 
\begin{equation*}
    -\left(\frac{2}{\eta}\Sigma^{-1} \cdot \frac{\pd \hat{A}^\noneq}{\pd \Sigma^{-1}}\right)_{ij} = -\left(\frac1{\eta}\avg{f_i(x-\mu)_j} + d\ \delta_{ij}\right)
\end{equation*}
which is to be compared to the right hand side of \ref{eqn:SigmaInvDyn}. 
Finally, we define the tensor $\text{Sym}_{ijkl} = \frac1{2}\left(\delta_{ik}\delta_{jl} + \delta_{il}\delta_{jk}\right)$ which for any matrix $M_{ij}$ obeys 
\begin{equation*}
    \text{Sym}_{ijkl}M_{kl} = \text{sym}(M)_{ij}, 
\end{equation*}
and so we have
\begin{equation}\label{eqn:intermediate}
    -\frac{2}{\eta}\text{Sym}_{ijmn}\Sigma^{-1}_{mo}\frac{\pd \hat{A}^\noneq}{\pd \Sigma^{-1}_{on}} = \mathbb{F}^{\Sigma^{-1}}_{ijkl}\dot{\Sigma}^{-1}_{kl}.
\end{equation}

We would like to invert $\mathbb{F}^{\Sigma^{-1}}_{ijkl}$, however, using its definition, we see
\begin{equation*}
    \mathbb{F}^{\Sigma^{-1}}_{ijkl} = \frac1{2}\Sigma_{im}\Sigma_{jn}\text{Sym}_{mnkl},
\end{equation*}
so that it contains the non-trivial projection tensor $\text{Sym}$ which projects matrices onto their symmetric components. Thus, the dynamical equations are only specified for the symmetric portion of $\dot{\Sigma}^{-1}$. Since we have assumed at the outset that $\Sigma^{-1}$ is symmetric, we lose nothing by assuming that $\dot{\Sigma}^{-1}$ is also symmetric.
Hence, we can ignore the projection, 
\begin{equation*}
\mathbb{F}^{\Sigma^{-1}}_{ijkl}\dot{\Sigma}^{-1}_{kl} = \frac1{2}\Sigma_{im}\Sigma_{jn}\text{Sym}_{mnkl}\dot{\Sigma}^{-1}_{kl} = \frac1{2}\Sigma_{ik}\Sigma_{jl}\dot{\Sigma}^{-1}_{kl}
\end{equation*}
and apply $\left(\frac1{2}\Sigma_{ik}\Sigma_{jl}\right)^{-1} = 2\Sigma^{-1}_{ik}\Sigma^{-1}_{jl}$ to both sides of Eqn. \ref{eqn:intermediate} to get
\begin{equation*}
    \dot{\Sigma}_{ij}^{-1} = -\frac{4}{\eta}\Sigma^{-1}_{ik}\Sigma^{-1}_{jl}\text{Sym}_{klmn}\Sigma^{-1}_{mo}\frac{\pd \hat{A}^\noneq}{\pd \Sigma^{-1}_{on}}.
\end{equation*}
We make one final step to recover the gradient flow form. Since $\Sigma^{-1}$ is symmetric, so is $\frac{\pd \hat{A}^\noneq}{\pd \Sigma^{-1}}$. This means we could also write 
\begin{equation*}
    \dot{\Sigma}_{ij}^{-1} = -\frac{4}{\eta}\Sigma^{-1}_{ik}\Sigma^{-1}_{jl}\text{Sym}_{klmn}\Sigma^{-1}_{mo}\text{Sym}_{onxy}\frac{\pd \hat{A}^\noneq}{\pd \Sigma^{-1}_{xy}},
\end{equation*}
or 
\begin{equation*}
    \dot{\Sigma}^{-1}_{ij} = -\mathbb{M}(\Sigma^{-1})_{ijkl}\frac{\pd \hat{A}^\noneq}{\pd \Sigma_{kl}^{-1}}
\end{equation*}
where $\mathbb{M}(\Sigma^{-1})_{ijkl} \equiv \frac{4}{\eta}\Sigma^{-1}_{im}\Sigma^{-1}_{jn}\text{Sym}_{mnop}\Sigma^{-1}_{oq}\text{Sym}_{qpkl}$ is positive semidefinite. To see this, we note that for any matrix $M_{ij}$, we have 
\begin{align*} 
M_{ij}\mathbb{M}(\Sigma^{-1})_{ijkl}M_{kl} &= M_{ij}\frac{4}{\eta}\Sigma^{-1}_{ik}\Sigma^{-1}_{jl}\text{Sym}_{klmn}\Sigma^{-1}_{mo}\text{Sym}_{onxy} M_{xy} \\
&= \frac{4}{\eta}\tr(\Sigma^{-1} \text{sym}(M) \Sigma^{-1} \text{sym}(M) \Sigma^{-1}) \\
&= \frac{4}{\eta}\norm{L^T\ \text{sym}(M)\Sigma^{-1}}^2_F \geq 0
\end{align*}
where $L$ is the Cholesky factor of $\Sigma^{-1}$, and $\norm{A}^2_F = \tr(A^TA)$ is the square of the Frobenius norm. 
\par 
Putting everything together, we see that 
\begin{align*}
    \dot{\mu} &= -\frac1{\eta} \frac{\pd \hat{A}^\noneq}{\pd \mu} \\
    \dot{\Sigma}^{-1}_{ij} &= -\mathbb{M}(\Sigma^{-1})_{ijkl}\frac{\pd \hat{A}^\noneq}{\pd \Sigma^{-1}_{kl}} \\
\end{align*}
for a positive semi-definite tensor $\mathbb{M}$, and hence the dynamics satisfy the gradient flow structure,
and the total rate of entropy production is non-negative
\begin{align*}
    T\frac{\diff}{\diff t} \hat{S}^{\text{tot}} &= -\frac{\pd \hat{A}^\noneq}{\pd \mu_i}\dot{\mu_i} - \frac{\pd \hat{A}^\noneq}{\pd \Sigma^{-1}_{ij}}\dot{\Sigma}^{-1}_{ij} \\
    &= \frac1{\eta}\norm{\frac{\pd \hat{A}^\noneq}{\pd \mu}}^2 + \frac{\pd \hat{A}}{\pd \Sigma^{-1}_{ij}}\mathbb{M}(\Sigma^{-1})_{ijkl}\frac{\pd \hat{A}}{\pd \Sigma^{-1}_{kl}} \geq 0.
\end{align*}

\subsection{Details for single colloidal particle and a linear external force}
In this example, we consider a single colloidal particle being pulled by a linear external force.
The system is assumed to obey the stochastic differential equation 
\begin{equation*}
    \diff x(t) = -\frac{k}{\eta}(x(t) - \lambda(t))\,\diff t + \sqrt{2d}\,\diff b(t).
\end{equation*}
The associated Fokker-Planck equation is then given by 
\begin{equation*} 
    \frac{\pd p }{\pd t} = \frac1{\eta}\frac{\pd }{\pd x}\left( \frac{\pd e}{\pd x}\ p \right) + d \frac{\pd^2 p}{\pd x^2}
\end{equation*}
where $e(x,\lambda) = \frac{k}{2}(\lambda - x)^2$ is the interaction energy and $\lambda(t)$ is the external protocol. Since the solution to this pde is the density of an Ornstein-Uhlenbeck process, which is a normal distribution for all time, we make the trail ansatz of
\begin{equation*}
 p(x,t) \approx \hat{p}(x, \mu,\sigma) = \frac1{\sqrt{2\pi \sigma^2}}\exp(-\frac{(x - \mu)^2}{2\sigma^2}). 
\end{equation*}
Thus, $\alpha = (\mu,\sigma)$ will be our internal variables.
The microscopic entropy is 
\begin{equation*}
  \hat{s} =-k_B\log(\hat{p}) =  k_B\frac{(x - \mu)^2}{2\sigma^2} + \frac{k_B}{2}\log(2\pi \sigma^2),
\end{equation*}
and the resulting test functions are
\begin{equation*}
    \frac{\pd\, \hat{s}}{\pd\, \mu}  = -k_B\frac{(x - \mu)}{\sigma^2} \qquad\qquad
    \frac{\pd\, \hat{s}}{\pd\, \sigma} = -k_B\frac{(x - \mu)^2}{\sigma^3} + \frac{k_B}{\sigma},
\end{equation*}
which are proportional to the first and second order Hermite polynomials as functions of $z = (x-\mu)/\sigma$.
First, we compute left hand side of Eqn. \ref{eqn:alphaDyn}
\begin{equation*}
    \avg{\frac{\pd\, \hat{s}}{\pd\, \mu}\frac{\pd\, \hat{s}}{\pd\, \mu}}_{\hat{p}} = \frac{k_B^2}{\sigma^2}, \qquad\quad
     \avg{\frac{\pd\, \hat{s}}{\pd\, \mu}\frac{\pd\, \hat{s}}{\pd\, \sigma}}_{\hat{p}} = 0, \qquad\quad
    \avg{\frac{\pd\, \hat{s}}{\pd\, \sigma}\frac{\pd\, \hat{s}}{\pd\, \sigma}}_{\hat{p}} = \frac{2k_B^2}{\sigma^2}.
\end{equation*}
Next, we compute the operator terms of the dynamical equations
\begin{equation*}
    \avg{\fL^\dagger  \, \frac{\pd\, \hat{s}}{\pd\, \mu}}_{\hat{p}} = k_B\frac{k}{\eta}\frac{(\mu - \lambda)}{\sigma^2}, \qquad\qquad \avg{\fL^\dagger \, \frac{\pd\, \hat{s}}{\pd\, \sigma}}_{\hat{p}} = k_B\frac{k}{\eta}\frac{2}{\sigma} - k_B d \frac{2}{\sigma^3}
\end{equation*}
(recall: $\fL^\dagger\,\psi = -\frac1{\eta} \frac{\pd e}{\pd x} \frac{\pd \psi}{\pd x} + d\, \frac{\pd^2 \psi}{\pd x^2}$). 
The dynamical equations are then
\begin{equation*}
    \dot{\mu} = -\frac{k}{\eta}(\mu - \lambda)\qquad\qquad
    \dot{\sigma} = -\frac{k}{\eta}\sigma\left(1 - \frac1{k\beta\sigma^2}\right)
\end{equation*}
which exactly recovers the dynamics for the mean and variance of the Ornstein-Uhlenbeck process.
\par
Next, we compute the thermodynamic quantities. First, the microscopic non-equilibrium free energy 
\begin{equation*}
    \hat{a}^{\noneq} = e - T\hat{s}
    =  \frac{k}{2}(\lambda - x)^2 -\frac1{\beta}\left(\frac{(x - \mu)^2}{2\sigma^2} + \frac1{2}\log(2\pi \sigma^2) \right)
\end{equation*}
and its expectation to get the macroscopic non-equilibrium free energy
\begin{equation*}
    \hat{A}^{\noneq} = \frac{k}{2}\left((\lambda - \mu)^2 + \sigma^2 \right)-\frac1{2\beta}\left(1 + \log(2\pi \sigma^2)\right).
\end{equation*}
Simple calculations show that the derivatives of the macroscopic non-equilibrium free energy give the dynamical equations, external force, and rate of total entropy production
\begin{align*}
    \dot{\mu} &= -\frac{k}{\eta}(\mu - \lambda) = -\frac1{\eta} \frac{\pd \hat{A}^{\noneq}}{\pd \mu} \\
    \nonumber \\
    \dot{\sigma} &= -\frac{k}{\eta}\sigma\left(1 - \frac1{k\beta \sigma^2} \right) = -\frac1{\eta} \frac{\pd \hat{A}^{\noneq}}{\pd \sigma} \\
    \nonumber\\ 
    \frac{d}{dt}\hat{W} &= \bbEb{\frac{\pd e}{\pd \lambda}}\dot{\lambda}= k(\lambda - \mu)\dot{\lambda} = \frac{\pd \hat{A}^{\noneq}}{\pd \lambda}\dot{\lambda} \\
    \nonumber \\
    T\frac{d}{dt}\hat{S}^{\text{tot}} &= - \frac{\pd \hat{A}^{\noneq}}{\pd \mu}\dot{\mu} - \frac{\pd \hat{A}^{\noneq}}{\pd \sigma} \dot{\sigma} =  \eta \dot{\mu}^2 + \eta \dot{\sigma}^2.
\end{align*}
Thus, as expected, we see that the motion of the internal variables obeys a gradient flow with respect to the macroscopic non-equilibrium free energy, and that the rate of total entropy production is always non-negative. 

\subsection{Details for a colloidal mass-spring-chain with double well interactions} 
Here we give all the calculations for the colloidal mass-spring-chain system with double-well interactions. Throughout this only this section, we will not use the index summation convention, but will rather explicitly write out all sums. We assume the system is composed of $N$ internal masses, the first being attached to a spring fixed at the origin, and the final being attached via a spring to the external protocol, $\lambda(t)$. 
In this section, all ``springs'' are assumed to be double-well as described below. 
As before, we assume the density of states obeys the Fokker-Planck equation 
\begin{equation*}
    \frac{\pd p}{\pd t} = \frac1{\eta}\frac{\pd}{\pd x} \cdot \left( \frac{\pd e}{\pd x}\ p \right) + d \Laplace_x \, p
\end{equation*}
where $\Laplace_x = \sum_{i=1}^N \frac{\pd^2}{\pd x_i^2}$ is the Laplacian and where the internal energy $e$ is given as follows. We assume the interaction energy of a single spring as a function of the spring length $z$ is given by the equation 
\begin{equation*}
 u(z) = \begin{cases} \frac{k_1}{2}(z + l_1)^2 & z \leq 0 \\
 \frac{k_2}{2}(z - l_2)^2 + h_2 & z > 0, \\ \end{cases}
\end{equation*}
where $h_2$ is chosen so that $u$ is continuous. The total energy is then $e(x,\lambda) = \sum_{i=1}^{N+1} u(x_i - x_{i-1})$ where here and throughout this section we define $x_0 \equiv 0$ and $x_{N+1} \equiv \lambda$ for notational simplicity. The force on mass $i= 1,\ldots,N$ is $f_i(x,\lambda) = -u'(x_i - x_{i-1}) + u'(x_{i+1} - x_i)$.

Although the internal energy is no longer quadratic in the positions of the masses, we approximate their probability density using a multivariate normal. We no longer expect to capture the density exactly. However, the approximation turns out to allow for an accurate prediction of the thermodynamic quantities, and the motion of a propagating phase front. We parameterize it using the mean $\mu$, and the inverse covariance matrix $\Sigma^{-1}$. Thus, we write
\begin{equation*}
    p(x,t\mid \lambda) \approx \hat{p}(x,\mu,\Sigma^{-1}) = \sqrt{\det( \frac1{2\pi}\Sigma^{-1})}\exp(-\frac1{2}(x - \mu)^T \Sigma^{-1} (x - \mu)),
\end{equation*}
and $\mu$ and $\Sigma^{-1}$ are interpreted as internal variables (i.e., $\alpha = (\mu,\Sigma^{-1})$ with $\Sigma^{-1}$ assumed to be symmetric). 
Much of what we need has been computed above in section \ref{sec:gradFlow}. To specify the dynamical equations for this model, we only need to compute $\avg{f_i}_{\hat{p}}$ to solve for $\dot{\mu}$ and $\avg{f_i(x-\mu)_j}_{\hat{p}}$ to solve for $\dot{\Sigma}^{-1}$.
\par 
As in the main text, we define the following notation. Let $D_1$ denote the $N$ by $N$, first order backwards difference matrix with zero boundary condition (i.e., $(D_1 x)_i = x_i - x_{i-1}$ for $i=2,\ldots,N$ and $(D_1x)_1 = x_1$). Hence, $(-D_1^T)$ is the $N$ by $N$ first order forwards difference matrix with zero boundary conditions (i.e., $(-D^Tx)_i = x_{i+1} - x_i$ for $i = 1,\ldots,N-1$ and $(-D^Tx)_N = -x_N$).
\par 
Let $\tau_i \equiv \sqrt{(D_1 \Sigma D_1^T)_{ii}}$, and $\tau_{N+1} \equiv \sqrt{\Sigma_{NN}}$. Define , for notational convenience, $(D_1x)_{N+1} \equiv \lambda - x_N$, $(D_1\mu)_{N+1} \equiv \lambda - \mu_N$, and $(D_1\Sigma)_{N+1,i} = -\Sigma_{N,i}$ for $i = 1,\ldots,N$ and for all $i=1,\ldots,N+1$ let $\hat{\Phi}_i \equiv \Phi((D_1\mu)_i/\tau_i)$, and $\hat{\phi}_i = \phi((D_1\mu)_i/\tau_i)$.
Since we assume $x$ is Gaussian under $\hat{p}$, $D_1x$ is also Gaussian under $\hat{p}$ with mean $D_1 \mu$ and covariance $D_1 \Sigma D_1^T$. Thus, we find 
\begin{align*}
    \bbP_{\hat{p}}(D_1x_i > 0) &= \int_0^\infty \phi\left(\frac{\xi - D_1\mu_i}{\tau_i}\right) \, \frac{\diff \xi}{\tau_i} \\
    &= \int_{-\frac{D_1\mu_i}{\tau_i}}^\infty \phi(\zeta)\,\diff \zeta \qquad \qquad \zeta = \frac{\xi - D_1\mu_i}{\tau_i} \\
    &= \int^{\frac{D_1\mu_i}{\tau_i}}_{-\infty} \phi(\zeta)\,\diff \zeta \qquad \qquad \phi(\zeta) = \phi(-\zeta) \\
    &= \Phi\left(\frac{D_1 \mu_i}{\tau_i}\right) \equiv \hat{\Phi}_i
\end{align*}
that $\hat{\Phi}_i$ is the predicted probability that spring $i$ is in the right minima. 
\par
Using the Gaussian integral identities presented in section \ref{sec:gaussInt}, we have 
\begin{align*}
    \avg{f_i}_{\hat{p}} &= \avg{u'(x_{i+1} - x_i) - u'(x_i - x_{i-1})}_{\hat{p}} \\  
    &= k_1\left((D_1\mu)_{i+1} + l_1\right)\left(1 - \Phi\left(\frac{(D_1\mu)_{i+1}}{\tau_{i+1}}\right)\right) + k_2\left((D_1\mu)_{i+1} - l_2\right) \Phi\left(\frac{(D_1\mu)_{i+1}}{\tau_{i+1}}\right) + (k_2 - k_1)\tau_{i+1}\phi\left(\frac{(D_1\mu)_{i+1}}{\tau_{i+1}}\right)\\
    &\quad -\left\{k_1\left((D_1\mu)_i + l_1\right)\left(1 - \Phi\left(\frac{(D_1\mu)_i}{\tau_i}\right)\right) + k_2\left((D_1\mu)_i - l_2\right) \Phi\left(\frac{(D_1\mu)_i}{\tau_i}\right) + (k_2 - k_1)\tau_i\phi\left(\frac{(D_1\mu)_i}{\tau_i}\right) \right\}\\
    &= \left(-D_1^Tm^\mu\right)_i + \delta_{i,N}m^\mu_{N+1}\qquad i = 1,\ldots,N\\ 
    \\
    \avg{f_i(x-\mu)_j}_{\hat{p}} &= \avg{u'(x_{i+1} - x_i)(x - \mu)_j}_{\hat{p}} - \avg{u'(x_i - x_{i-1})(x-\mu)_j}_{\hat{p}} \\ 
    & =(D_1\Sigma)_{i+1,j}\left(k_1\left(1-\Phi\left(\frac{(D_1\mu)_{i+1}}{\tau_{i+1}}\right)\right) + k_2\Phi\left(\frac{(D_1\mu)_{i+1}}{\tau_{i+1}}\right)-(k_1l_1 + k_2l_2)\frac{\phi\left(\frac{(D_1\mu)_{i+1}}{\tau_{i+1}}\right)}{\tau_{i+1}} \right) \\ 
    &\quad -\left\{(D_1\Sigma)_{i,j}\left(k_1\left(1-\Phi\left(\frac{(D_1\mu)_i}{\tau_i}\right)\right) + k_2\Phi\left(\frac{(D_1\mu)_i}{\tau_i}\right)-(k_1l_1 + k_2l_2)\frac{\phi\left(\frac{(D_1\mu)_i}{\tau_i}\right)}{\tau_i} \right)\right\} \\
    &= \eta (D^w_2\Sigma)_{ij} \qquad i,j = 1,\ldots,N
\end{align*}
where 
\begin{align*}
    m^\mu_i &= k_1\left((D_1\mu)_i + l_1\right)\left(1 - \Phi\left(\frac{(D_1\mu)_i}{\tau_i}\right)\right) + k_2\left((D_1\mu)_i - l_2\right) \Phi\left(\frac{(D_1\mu)_i}{\tau_i}\right) + (k_2 - k_1)\tau_i\phi\left(\frac{(D_1\mu)_i}{\tau_i}\right) \\
    (D^w_2)_{ij} &= \sum_{k=1}^{N} (-D_1^T)_{ik}w_k(D_1)_{kj} - \delta_{i,N}\delta_{j,N}w_{N+1} \\
    w_i &= \frac{k_1}{\eta}\left(1-\Phi\left(\frac{(D_1\mu)_i}{\tau_i}\right)\right) + \frac{k_2}{\eta}\Phi\left(\frac{(D_1\mu)_i}{\tau_i}\right)-\frac1{\eta}(k_1l_1 + k_2l_2)\frac{\phi\left(\frac{(D_1\mu)_i}{\tau_i}\right)}{\tau_i} \quad i = 1,\ldots,N+1.
\end{align*}
Plugging these into Eqn. \ref{eqn:MuDyn} and Eqn. \ref{eqn:SigmaInvDyn}, and inverting $\mathbb{F}^{\Sigma^{-1}}_{ijkl}$ (since the right hand side of Eqn. \ref{eqn:SigmaInvDyn} is symmetric), we get
\begin{align}
    \dot{\mu} &= \frac1{\eta}(-D^T_1m^\mu) + \frac1{\eta}m^\mu_{N+1}\bm{e}_N \nonumber\\ 
    \dot{\Sigma}^{-1} &= -2\left(\text{sym}(D_2^w\Sigma^{-1}) + d\, \Sigma^{-2}\right) \nonumber
\end{align}
as given in the main text. 
\par

Now we turn to computing the thermodynamic quantities using the non-equilibrium free energy, $\hat{A}^\noneq$. By definition, its formula is 
\begin{equation*}
    \hat{A}^{\noneq} = \hat{E} - T\hat{S} =  \avg{e}_{\hat{p}} + \frac1{\beta}\avg{\log(\hat{p})}_{\hat{p}} .
\end{equation*}
The second term is simply the negative of the entropy of a multivariate normal
\begin{equation*}
    \avg{\log(\hat{p})}_{\hat{p}} = -\frac{N}{2}(1 + \log(2\pi)) + \frac1{2}\log(\det(\Sigma^{-1})).
\end{equation*}
The approximate energy can be computed via the Gaussian integrals in section \ref{sec:gaussInt}. 
First, we compute the expected value of the interaction energy with respect to an arbitrary Gaussian random variable $X \sim N(\nu,\sigma^2)$  
\begin{align*}
    U(\nu,\sigma) \equiv \avg{u(X)}_{X\sim N(\nu,\sigma^2)}  &= \frac{k_1}{2}\left((\sigma^2 + (\nu + l_1)^2)\Phi(-\nu/\sigma) - \sigma(\nu + 2l_1)\phi(-\nu/\sigma) \right) \\
    &\quad + \frac{k_2}{2}\left((\sigma^2 + (\nu-l_2)^2)\Phi(\nu/\sigma) + \sigma(\nu - 2l_2)\phi(\nu/\sigma)  \right) + h_2\Phi(\nu/\sigma)
\end{align*}
and so 
\begin{equation*}
    \hat{E} = \avg{e}_{\hat{p}} = \sum_{i=1}^N U((D_1\mu)_i,\sqrt{(D_1\Sigma D_1^T)_{ii}}) + U(\lambda - \mu_N,\sqrt{\Sigma_{NN}}) = \sum_{i=1}^{N+1} U((D_1\mu)_i,\tau_i) 
\end{equation*}
and 
\begin{equation*}
    \hat{A}^\noneq = \hat{E} - T\hat{S} = \sum_{i=1}^{N+1} U((D_1\mu)_i,\tau_i)  - \frac{N}{2\beta}(1 + \log(2\pi)) + \frac1{2\beta}\log(\det(\Sigma^{-1})).
\end{equation*}
Plugging these into equation for $\hat{A}^\noneq$ and taking derivatives (note $\frac{\pd U}{\pd \nu}$ and $\frac{\pd U}{\pd \sigma}$ are both computed in section \ref{sec:gaussInt}), we see
\begin{align*}
    \frac{\pd \hat{A}^{\noneq}}{\pd \mu_j} &= (D_1^T)_{ji} \frac{\pd U}{\pd \nu}(D_1 \mu_i,\tau_i) - \delta_{jN}\frac{\pd U}{\pd \nu}(\lambda - \mu_N,\tau_{N+1}) \\
    &= -\eta \dot{\mu}_j \\
    \\
    \frac{\pd \hat{A}^{\noneq}}{\pd \lambda} &= \frac{\pd U}{\pd \nu}(\lambda - \mu_N,\sqrt{\Sigma_{NN}}) = m_{N+1}^\mu \\
    \\
    \frac{\pd \hat{A}^{\noneq}}{\pd \Sigma^{-1}_{ij}} &= \sum_{k=1}^{N+1} \frac{\pd U}{\pd \sigma}(D\mu_k,\tau_k)\frac{\pd \tau_k}{\pd \Sigma_{ij}^{-1}} + \frac1{2\beta}\Sigma_{ij}\\
    \\
    &=  \sum_{k=1}^N\frac{\eta}{2}\Sigma_{il}(-D_1^T)_{lk}w_k(D_1)_{km}\Sigma_{mj} - \frac{\eta}{2}\Sigma_{iN} w_{N+1} \Sigma_{Nj} + \frac1{2\beta}\Sigma_{ij} \\
    \\
    &= \frac{\eta}{2}\left(\Sigma D_2^w \Sigma + d\ \Sigma\right)_{ij}.
\end{align*}
It's a straight forward calculation to verify that the gradient flow structure holds as proved in section \ref{sec:gradFlow} for this specific example. The fluxes of the internal variables are related to their affinities by 
\begin{align*}
    \dot{\mu} &= -\frac1{\eta}\frac{\pd \hat{A}^\noneq}{\pd \mu} \\
    \dot{\Sigma}^{-1} &= -\mathbb{M}(\Sigma^{-1}):\frac{\pd \hat{A}^\noneq}{\pd \Sigma^{-1}}, \\
\end{align*}
where, we recall that 
\begin{equation*}
    \mathbb{M}(\Sigma^{-1})_{ijkl} = \sum_{m,n,o,p,q=1}^N\frac{4}{\eta}\Sigma^{-1}_{im}\Sigma^{-1}_{jn}\text{Sym}_{mnop}\Sigma^{-1}_{oq}\text{Sym}_{qpkl} = \frac1{\eta}\left(\Sigma^{-2}_{ik}\Sigma^{-1}_{jl} + \Sigma^{-1}_{ik}\Sigma^{-2}_{jl} + \Sigma^{-2}_{il}\Sigma^{-1}_{jk} + \Sigma^{-1}_{il}\Sigma^{-2}_{jk} \right)
\end{equation*} 
is positive semi-definite. Thus, the rate of total entropy production is non-negative. 
\subsection{Continuum limit of the colloidal, double-well mass-spring-chain}
We now derive the dynamical equations for the internal variables in the continuum limit of infinitely many masses rescaled to a finite system size. In order to ensure a finite energy and dissipation in the limit, the parameters of the interaction energy, namely $l_1$, $l_2$, $k_1$, $k_2$, and $h_2$, along with viscosity, $\eta$, and inverse temperature, $\beta$, must be correctly rescaled. We allow for these values to depend on the number of masses $N$, and denote this dependence on $N$ via a superscript, e.g., $k_1^{(N)}$. To motivate how this scaling should take place, we first consider a mass-spring-chain with quadratic interaction energy 
\begin{equation*}
    u^{(N)}(z) = \frac{k^{(N)}}{2}(z - l^{(N)})^2.
\end{equation*}
The total energy of the $N$-mass system, assuming the right most mass is attached to an external control, fixed at a location of $\lambda$ is given by 
\begin{align*}
    e^{(N)}(x^{(N)},\lambda) &= u^{(N)}(x_1^{(N)}) + \sum_{i=2}^N u^{(N)}(x_i^{(N)} - x_{i-1}^{(N)}) + u^{(N)}(\lambda - x_N^{(N)}) = \sum_{i=1}^{N+1} u^{(N)}(x_i^{(N)} - x_{i-1}^{(N)}) \\
    &= \frac{k^{(N)}}{2}\left\{ (D^{(N)}_1x^{(N)})^T (\text{Id}^{(N)} + \bm{1}^{(N)}(\bm{1}^{(N)})^T)D^{(N)}_1x^{(N)} - 2\lambda (\bm{1}^{(N)})^TD^{(N)}_1x^{(N)} + (N+1)(l^{(N)})^2 - 2\lambda l^{(N)} + \lambda^2 \right\} \\
    &=\frac{k^{(N)}}{2}\left\{ (D^{(N)}_1x^{(N)} - \frac{\lambda}{N+1}\bm{1}^{(N)})^T (\text{Id}^{(N)} + \bm{1}^{(N)}(\bm{1}^{(N)})^T)(D^{(N)}_1x^{(N)} - \frac{\lambda}{N+1}\bm{1}^{(N)})  + (N+1)\left(l^{(N)} - \frac{\lambda}{N+1}\right)^2 \right\}
\end{align*}
where we have defined $x_0 \equiv 0$, $x_{N+1} \equiv \lambda$, $D^{(N)}_1$ as the $N$ by $N$ first order backward difference as above, $\text{Id}^{(N)}$ to be the $N$ by $N$ identity matrix, and $\bm{1}^{(N)} = \mat{1 & 1 & \ldots & 1}^T \in \bb{R}^N$. 
\par 
Since temperature is an intensive quantity, it must be constant, $T^{(N)} = T^0$. The equilibrium density given by 
\begin{equation*}
    p_\text{eq}(x^{(N)}\mid \lambda) = \frac{\exp(-\beta^{(N)} e^{(N)}(x^{(N)},\lambda))}{Z(N)}
\end{equation*}
is Gaussian for any fixed $\lambda$, with mean 
\begin{equation*}
    (D_1^{(N)})^{-1} \frac{\lambda}{N+1} \bm{1}^{(N)} = \frac{\lambda}{N+1} \mat{1 & 2 & \cdots & N}^T \equiv \nu 
\end{equation*} 
and covariance 
\begin{equation*}
\frac1{k^{(N)}\beta^{(N)}} \left((D_1^{(N)})^T\left(\text{Id}^{(N)} + \bm{1}^{(N)}(\bm{1}^{(N)})^T\right)D_1^{(N)}\right)^{-1} \equiv \Xi^{-1}. 
\end{equation*}
We see that the energy writes 
\begin{equation*}
    e^{(N)} = \frac{1}{2\beta}(x^{(N)} - \nu)^T\Xi^{-1}(x^{(N)} - \nu) + (N+1)\frac{k^{(N)}}{2}\left(l^{(N)} - \frac{\lambda}{N+1}\right)^2
\end{equation*}
and hence the average energy is
\begin{equation}
    E^{(N)} = \avg{e^{(N)}}_{p_\text{eq}} = \frac{N}{2\beta^{(N)}} + (N+1)\frac{k^{(N)}}{2}\left(l^{(N)} - \frac{\lambda}{N+1}\right)^2. \label{eqn:energy}
\end{equation}
First, for fixed $\lambda \equiv (N+1)l^{(N)}$, we see that for the energy due to thermal vibrations to be finite, we need $\beta^{(N)} = N\beta^0$. Since the bath temperature must be constant, the effective Boltzmann constant $k_B^{(N)}$ must scale like $k_B^{(N)} = k_B^0/N$. The second term in \ref{eqn:energy} gives the mechanical energy. In order for this term to have zero energy at the finite extension $\lambda = l^0$, we need $l^{(N)} = l^0/(N+1)$. The mechanical energy at arbitrary extension $\lambda$ becomes
\begin{equation*}
    E^{(N)}_\text{mech} = \frac{k^{(N)}}{2(N+1)}(l^0 - \lambda)^2.
\end{equation*}
Hence, $k^{(N)} = (N+1)k^0$ is required for finite mechanical energy for all $\lambda$. 
Finally, to determine $\eta^{(N)}$ we consider a quasi-static pulling in which the external control slowly changes by a small $\diff\lambda$ over a small time $\diff t$ for a system at zero temperature. Since the system always remains in the minimum energy configuration, we know that each spring stretches by a length $\frac{\diff \lambda}{N+1}$ over the same time $\diff t$. Or, $\diff (x_{i}^{(N)} - x_{i-1}^{(N)}) = \frac{\diff \lambda}{N+1}$ for all $i=1,\ldots,N+1$. Hence, we know that each mass must move by $\diff x^{(N)}_i = \frac{i}{N+1}\diff \lambda$ over the time $\diff t$. In order to achieve an energy dissipation rate due to the viscosity of the fluid surrounding the system which is finite, we need
\begin{equation*}
    \dot{D}(\diff\lambda,\diff t) = \sum_{i=1}^N \eta^{(N)}\left(\frac{\diff x_i^{(N)}}{\diff t}\right)^2  = \eta^{(N)}\left(\frac{\diff \lambda}{\diff t}\right)^2 \frac1{(N+1)^2}\sum_{i=1}^N i^2 = \eta^{(N)}\left(\frac{\diff \lambda}{\diff t}\right)^2 \frac{N(N+1)(2N+1)}{6(N+1)^2}
\end{equation*}
to have a finite limit.
Although, taking $\eta^{(N)} = \eta^0\frac{6(N+1)^2}{N(N+1)(2N+1)}$ leads to a complete lack of dependence on $N$, we simply consider $\eta^{(N)} = \frac{\eta^0}{(N+1)}$ which suffices to produce a finite energy dissipation rate in the continuum limit.
\par 
Returning to the double-well mass-spring-chain system, we are motivated to use $l_1^{(N)} = l_1^0/(N+1)$, $l_2^{(N)} = l_2^0/(N+1)$, $k_1^{(N)} = k_1^0(N+1)$, $k_2^{(N)} = k_2^0(N+1)$, $\eta^{(N)} = \eta^0/(N+1)$ and, $\beta^{(N)} = (N+1)\beta^0$ for our parameters in the interaction energy, viscosity, and temperature ($h_2^{(N)} = \frac{k_1^{(N)}}{2}(l_1^{(N)})^2 - \frac{k_2^{(N)}}{2}(l_2^{(N)})^2$ is still used to ensure the interaction potential is continuous). 
For notational, we first define 
\begin{align*}
    \hat{\Phi}_k^{(N)} &= \Phi\left(\frac{(D_1^{(N)}\mu^{(N)})_k}{\sqrt{\left(D_1^{(N)} \Sigma^{(N)} (D_1^{(N)})^T\right)_{kk}}}\right) \\
    \hat{\phi}_k^{(N)} &= \phi\left(\frac{(D_1^{(N)}\mu^{(N)})_k}{\sqrt{\left(D_1^{(N)} \Sigma^{(N)} (D_1^{(N)})^T\right)_{kk}}}\right). 
\end{align*}
The resulting dynamical equations for $\mu^{(N)}_i$ for $i=1,\ldots,N-1$ are given by (ignoring $i = N$ and the term with $m^\mu_{N+1}$ for now)
\begin{align*}
    \dot{\mu}^{(N)}_i &= - \sum_{j=1}^N \frac1{\eta^{(N)}}(D_1^{(N)})^T_{ij}\bigg\{k_1^{(N)}\left((D_1^{(N)}\mu^{(N)})_j + l_1^{(N)}\right)\left(1 - \hat{\Phi}_j^{(N)}\right)\\
    &\qquad\qquad+ k_2^{(N)}\left((D_1^{(N)}\mu^{(N)})_j - l_2^{(N)}\right)\hat{\Phi}_j^{(N)} \\
    &\qquad\qquad+ (k_2^{(N)} - k_1^{(N)})\sqrt{(D_1^{(N)}\Sigma^{(N)}(D_1^{(N)})^T)_{jj}}\,\hat{\phi}^{(N)}_j\bigg\} \\
    \\
 \dot{\mu}^{(N)}_i &= - \sum_{j=1}^N \frac{N+1}{\eta^0}(D_1^{(N)})^T_{ij}\bigg\{k_1^0(N+1)\left((D_1^{(N)}\mu^{(N)})_j + \frac{l_1^0}{(N+1)}\right)\left(1 - \hat{\Phi}_j^{(N)}\right)\\
    &\qquad\qquad+ k_2^0(N+1)\left((D_1^{(N)}\mu^{(N)})_j - \frac{l_2^0}{(N+1)}\right)\hat{\Phi}_j^{(N)} \\
    &\qquad\qquad+ (k_2^0 - k_1^0)(N+1)\sqrt{(D_1^{(N)}\Sigma^{(N)}(D_1^{(N)})^T)_{jj}}\,\hat{\phi}^{(N)}_j\bigg\}. \\
    \\
\end{align*}
As it stands, the reference frame for the internal variables are $i,j \in 1,\ldots,N+1$. In order to use the unit interval and square as reference frame for $\mu$ and $\Sigma^{-1}$ respectively, we simply employ the mapping $\mu^{(N)}(i/(N+1)) = \mu^{(N)}_i$ and $(\Sigma^{(N)})^{-1}(i/(N+1),j/(N+1)) = (\Sigma^{(N)})^{-1}_{ij}$. Thus, we see that $\sum_{j=1}^N(N+1)(D_1)_{ij}$, the backwards finite difference, converges to a derivative with respect to the coordinate corresponding to the $j^\text{th}$ index. Likewise $\sum_{j=1}^N(N+1)(-D_1^T)_{ij}$, the forward finite difference, converges to a derivative with respect to the coordinate corresponding to the $j^\text{th}$ index as well. Hence, by multiplying and dividing by $(N+1)$ in the numerator and denominator of $\hat{\Phi}^{(N)}$ and $\hat{\phi}^{(N)}$, we see that 
\begin{align*}
    \lim_{N\rightarrow \infty} \hat{\Phi}^{(N)} &= \Phi\left(\frac{\frac{\pd \mu}{\pd x}(x,t)}{\sqrt{\frac{\pd^2\Sigma}{\pd x\pd y}(x,x,t)}}\right) \equiv \hat{\Phi}(x,t) \\
    \lim_{N\rightarrow \infty} \hat{\phi}^{(N)} &= \phi\left(\frac{\frac{\pd \mu}{\pd x}(x,t)}{\sqrt{\frac{\pd^2\Sigma}{\pd x\pd y}(x,x,t)}}\right) \equiv \hat{\phi}(x,t). 
\end{align*}

By multiplying and dividing by $(N+1)$ inside $\Phi$ and $\phi$ we see that the limiting equation must be 
\begin{align*}
    \dot{\mu}(x,t) &= \frac1{\eta^{0}}\frac{\pd}{\pd x}\bigg\{k_1^{0}\left(\frac{\pd\mu}{\pd x}(x,t) + l_1^{0}\right)\left(1 - \hat{\Phi}(x,t)\right)\\
    &\qquad\qquad+ k_2^{0}\left(\frac{\pd \mu}{\pd x}(x,t) - l_2^{0}\right)\hat{\Phi}(x,t) \\
    &\qquad\qquad+ (k_2^{0} - k_1^{0})\sqrt{\frac{\pd^2\Sigma}{\pd x\pd y}(x,x,t)}\,\hat{\phi}(x,t)\bigg\}.
\end{align*}
We see then that the $m^\mu_{N+1}$ simply encodes the interaction of the last mass with a fictitious $N+1^\text{st}$ mass with controlled position $\lambda$. Hence, the above equation is the full limiting equation for $\mu(x)$ and the boundary conditions are $\mu(0) = 0$ and $\mu(1) = \lambda$. 
\par
We carry out the exact same analysis to transform the equation for $(\Sigma^{(N)}_{ij})^{-1}$ into an equation for $\Sigma^{-1}(x,y,t)$ in the limit as $N\rightarrow \infty$. We begin with the matrix $D_2^w$.
Thus, we have for $i,j = 1,\ldots,N-1$ (again, ignoring the final term)
\begin{align*}\hspace{-.6 in}
    (D_2^w)^{(N)}_{ij} &= \sum_{k=1}^N (-D_1^{(N)})^T_{ik}\left(\frac{k_1^{(N)}}{\eta^{(N)}}(1 - \hat{\Phi}^{(N)}_k) + \frac{k_2^{(N)}}{\eta^{(N)}}\hat{\Phi}^{(N)}_k - \frac1{\eta^{(N)}}(k_1^{(N)}l_1^{(N)} +k_2^{(N)}l_2^{(N)})\frac{\hat{\phi}^{(N)}_k}{\sqrt{(D_1^{(N)}\Sigma^{(N)} (D_1^{(N)})^T)_{kk}}} \right)(D_1^{(N)})_{kj} \\
     &= \sum_{k=1}^N (-D_1^{(N)})^T_{ik}\bigg(\frac{(N+1)^2 k_1^{0}}{\eta^{0}}(1 - \hat{\Phi}^{(N)}_k) + \frac{(N+1)^2 k_2^{0}}{\eta^{0}}\hat{\Phi}^{(N)}_k \\
     &\qquad - \frac{(N+1)}{\eta^{0}}(k_1^{0}l_1^{0} +k_2^{0}l_2^{0})\frac{(N+1)\hat{\phi}^{(N)}_k}{\sqrt{((N+1)D_1^{(N)}\Sigma^{(N)}(N+1) (D_1^{(N)})^T)_{kk}}} \bigg)(D_1^{(N)})_{kj} \\
      &= \sum_{k=1}^N (N+1)(-D_1^{(N)})^T_{ik}\left(\frac{k_1^{0}}{\eta^{0}}(1 - \hat{\Phi}^{(N)}_k) + \frac{k_2^{0}}{\eta^{0}}\hat{\Phi}^{(N)}_k - \frac1{\eta^{0}}(k_1^{0}l_1^{0} +k_2^{0}l_2^{0})\frac{\hat{\phi}^{(N)}_k}{\sqrt{((N+1)D_1^{(N)}\Sigma^{(N)}(N+1) (D_1^{(N)})^T)_{kk}}} \right)((N+1)D_1^{(N)})_{kj}.
\end{align*}
Since both $\hat{\Phi}^{(N)}$ and $\hat{\phi}^{(N)}$ will converge as before, in the limit as $N\rightarrow \infty$, we get 
\begin{equation*}
    \lim_{N\rightarrow \infty} D_2^w = \frac{\pd}{\pd x} \left(\frac{k_1^{0}}{\eta^{0}}(1 - \hat{\Phi}(x,t)) + \frac{k_2^{0}}{\eta^{0}}\hat{\Phi}(x,t) - \frac1{\eta^{0}}(k_1^{0}l_1^{0} +k_2^{0}l_2^{0})\frac{\hat{\phi}(x,t)}{\sqrt{\frac{\pd^2\Sigma}{\pd x\pd y}(x,x,t)}} \right) \frac{\pd}{\pd x} \equiv  \frac{\pd}{\pd x} w(x,t) \frac{\pd}{\pd x} \qquad x \in  (0,1).
\end{equation*}
As with the equation for $\mu$, we examine the first and last indices to reveal the boundary conditions. The boundary term, $(D_2^w)^{(N+1)}_{N,N}$, contains the extra $w_{N+1}$. The form of $w_{N+1}$ is the same as before expect for the fact that $\hat{\Phi}^{(N)}_{N+1}$ and $\hat{\phi}^{(N)}_{N+1}$ are defined through 
\begin{equation*}
    \hat{\Phi}^{(N)}_{N+1} = \Phi\left(\frac{\lambda - \mu_N}{\sqrt{\Sigma_{NN}}} \right)
\end{equation*}
and likewise for $\hat{\phi}_{N+1}^{(N)}$. As before, $(N+1)(\lambda - \mu_N) \rightarrow \frac{\pd \mu}{\pd x}(1,t)$ as $N\rightarrow\infty$ assuming the boundary condition of $\mu(1,t) = \lambda$. Likewise, $(N+1)^2\Sigma_{NN} \rightarrow \frac{\pd^2\Sigma}{\pd x\pd y}(1,1,t)$ as $N \rightarrow \infty$ assuming the boundary condition $\Sigma(1,y,t) = \Sigma(x,1,t) = 0$ for all $x,y\in[0,1]$. The variance of the $N+1$st spring would be $\Sigma_{N+1,N+1} - \Sigma_{N+1,N} - \Sigma_{N,N+1} + \Sigma_{N,N}$ where $\Sigma_{N+1,N+1}$ is the variance of the fictitious $N+1$st mass. However, since this fictitious mass's location is held fixed by the external control, $\Sigma_{N+1,N+1} = \Sigma_{N+1,i} = \Sigma_{i,N+1} \equiv 0$ for all $i\in 1,\ldots,N$. Hence the analogous boundary condition in the continuum limit. Using the same logic, noticing that the variance of the first spring is $(D_1^{(N+1)}\Sigma^{(N)}(D_1^{(N)})^T)_{11} = \Sigma_{11} = \Sigma_{11} - \Sigma_{1,0} - \Sigma_{0,1} + \Sigma_{00}$ assuming $\Sigma_{00} = \Sigma_{0,i} = \Sigma_{i,0} = 0$ for all $i = 1,\ldots,N$ reveals the further boundary condition of $\Sigma(0,y,t) = \Sigma(x,0,t) = 0$ for all $x,y\in[0,1]$. 
\par
Now we turn to the rest of the equation for $(\Sigma^{(N)})^{-1}$
\begin{equation}\label{eqn:SigmaInvDyn1}
    \frac{\pd}{\pd t} (\Sigma^{(N)})^{-1} = -2\left(\text{sym}\left(D_2^w (\Sigma^{(N)})^{-1}\right) + d^{(N)}(\Sigma^{(N)})^{-2}\right).
\end{equation}
In order avoid difficulties of inverting the continuum function $\Sigma^{-1}(x,y)$, we will now instead use Eq. \ref{eqn:SigmaInvDyn1} to derive the dynamics of $\Sigma^{(N)}$.
Since we know 
\begin{equation*}
  \Sigma^{(N)}(\Sigma^{(N)})^{-1} = \mathbb{I}  
\end{equation*}
for all time, we may differentiate in time to get 
\begin{equation*}
  \dot{\Sigma}^{(N)}(\Sigma^{(N)})^{-1} + \Sigma^{(N)}(\dot{\Sigma}^{(N)})^{-1}= 0.
\end{equation*}
Moving the second term to the right hand side and multiplying by $\Sigma^{(N)}$ on the right gives
\begin{equation*}
  \dot{\Sigma}^{(N)} = - \Sigma^{(N)}(\dot{\Sigma}^{(N)})^{-1} \Sigma^{(N)}.
\end{equation*}
As $D_2^w$ is symmetric
\begin{equation*}
    \Sigma^{(N)}\text{sym}\left(D_2^w(\Sigma^{(N)})^{-1}\right)\Sigma^{(N)} = \text{sym}\left(D_2^w \Sigma^{(N)}\right),
\end{equation*}
we finally get 
\begin{equation}\label{eqn:SigmaDyn}
    \dot{\Sigma}^{(N)} = 2\left(\text{sym}\left(D_2^w\Sigma^{(N)}\right) + d^{(N)}\text{Id}\right),
\end{equation}
where
\begin{align*}
 \text{sym}\left(D_2^w \Sigma^{(N)}\right) &= \frac1{2}\sum_{k=1}^N (D_2^w)_{ik} (\Sigma^{(N)})_{kj} +  (\Sigma^{(N)})_{ik}(D_2^w)_{kj} \\
 &= \frac1{2}\sum_{k=1}^N (D_2^w)_{ik} (\Sigma^{(N)})_{kj} +  (D_2^w)_{jk}(\Sigma^{(N)})_{ki}.
\end{align*}
Hence
\begin{equation*}
    \lim_{N\rightarrow \infty} \text{sym}\left(D_2^w \Sigma^{(N)}\right) = \frac1{2}\left(\frac{\pd}{\pd x}w(x,t)\frac{\pd}{\pd x} + \frac{\pd}{\pd y}w(y,t)\frac{\pd}{\pd y} \right) \Sigma(x,y,t) \equiv \Laplace^w(x,y,t)\Sigma(x,y,t).
\end{equation*}
On the other hand, we have $d^{(N)} \equiv \frac1{\beta^{(N)}\eta^{(N)}} = \frac{(N+1)}{(N+1)\beta^0\eta^0} = d^0$.
Finally, since it is unclear how the identity matrix, $\text{Id}$, should scale to the continuum limit, we consider the scaling limit of the weak form of the equation. Let $\xi,\psi:[0,1]\rightarrow \bb{R}$ be smooth functions and define for each $N$, $\xi^{(N)} \equiv \mat{\xi_1^{(N)}&\cdots&\xi_N^{(N)}}^T$ and $\psi^{(N)} \equiv \mat{\psi_1^{(N)} & \cdots & \psi_N^{(N)}}^T$ with $\xi_i^{(N)} = \xi(i/N)$ and $\psi_i^{(N)} = \psi(i/N)$ for all $i=1\ldots N$. Then we can compute 
\begin{align*}
    \lim_{N\rightarrow \infty} \sum_{i,j=1}^N \dot{\Sigma}^{(N)}_{ij}\xi_i^{(N)}\psi_j^{(N)} \frac1{N^2} &= \int_0^1\int_0^1 \dot{\Sigma}(x,y,t)\xi(x)\psi(y)\diff x\diff y \\
    \lim_{N\rightarrow \infty} \sum_{i,j=1}^N \text{sym}\left(D_2^w\Sigma^{(N)}\right)_{ij}\xi_i^{(N)}\psi_j^{(N)} \frac1{N^2}&= \int_0^1\int_0^1 (\Laplace^w \Sigma)(x,y,t)\xi(x)\psi(y) \diff x\diff y \\
    \lim_{N\rightarrow \infty} \sum_{i,j=1}^N d^{(N)}\delta_{ij}\xi_i^{(N)}\psi_j^{(N)} \frac1{N^2} &= \lim_{N\rightarrow \infty} \frac{d^0}{N} \sum_i \xi_i^{(N)}\psi_i^{(N)} \frac1{N} = 0,
\end{align*}
where the last term vanishes because only one factor of $1/N$ is needed in the limit of the integral, $\sum_i \xi_i^{(N)}\psi_i^{(N)} \frac1{N} \rightarrow \int_0^1 \xi(x)\psi(x) \diff x$. Thus, the partial differential equation for $\Sigma$ as $N \rightarrow \infty$ is simply
\begin{equation*}
    \dot{\Sigma} = 2\Laplace^w \Sigma
\end{equation*}
which has the form of a modified heat equation. As described above, the associated boundary conditions are
\begin{align*}
    \Sigma(0,y,t) = \Sigma(x,0,t) = \Sigma(x,1,t) = \Sigma(1,y,t) = 0
\end{align*}
in direct analogy to the fictitious $0$th and $N+1$st held fixed at the origin and by the external control respectively in the discrete system. 
\par
As a final point, we note that the alternative choice of requiring $\beta^{(N)} = \beta^0$ leads to an equation for $\dot{\Sigma}$ in which the diffusion coefficient does not vanish
\begin{equation*}
\dot{\Sigma} = 2\Laplace^w \Sigma + d^0 \delta(x-y)
\end{equation*}
where $\delta(x-y)$ is the Dirac delta function and $d^0 = 1/\beta^0\eta^0$ as before. We do not pursue this direction because energy due to thermal fluctuations in the continuum limit is infinite. Moreover, the system contains finite sized fluctuations relative to the system size, and non-deterministic behavior in the continuum limit. 
\par 
Now, we turn to interpreting the continuum equation. We begin by defining two new term. Since $\mu(x,t)$ is the average displacement of the system, 
\begin{equation*}
    \epsilon(x,t) \equiv \frac{\pd\mu}{\pd x}(x,t)
\end{equation*}
defines the average strain. Likewise, $\Sigma(x,y,t)$ is the covariance of the displacement, so we define 
\begin{equation*}
    \E(x,y,t) = \frac{\pd^2\Sigma}{\pd x\pd y}(x,y,t)
\end{equation*}
to be the covariance of the strain. We then see that the phase fraction can be expressed as a function of $\epsilon$ and $\E$
\begin{equation*}
    \hat{\Phi}(x,t) = \Phi\left( \frac{\frac{\pd\mu}{\pd x}(x,t)}{\sqrt{\frac{\pd^2\Sigma}{\pd x\pd y}(x,x,t)}} \right) = \Phi\left(\frac{\epsilon(x,t)}{\sqrt{\E(x,x,t)}}\right) \equiv \hat{\Phi}(\epsilon(x,t),\E(x,x,t)).
\end{equation*}
Moreover, we can make the identification
\begin{equation*}
    \frac{\hat{\phi}(x,t)}{\sqrt{\frac{\pd^2\Sigma}{\pd x\pd y}(x,x,t)}} = \frac{\phi\left(\frac{\frac{\pd\mu}{\pd x}(x,t)}{\sqrt{\frac{\pd^2\Sigma}{\pd x\pd y}(x,x,t)}}\right)}{\sqrt{\frac{\pd^2\Sigma}{\pd x\pd y}(x,x,t)}}  = \frac{\pd \hat{\Phi}}{\pd \epsilon}(\epsilon(x,t),\E(x,x,t)),
\end{equation*}
which allows us to write the continuum dynamical equations without $\hat{\phi}$ explicitly, and in a much simpler form
\begin{align*}
    \dot{\mu} &= \frac1{\eta^0}\frac{\pd}{\pd x}\left\{k_1^0(\epsilon + l_1^0)(1 - \hat{\Phi}) + k_2^0(\epsilon - l_2^0)\hat{\Phi} + (k_2^0 - k_1^0)\E(x,x,t)\frac{\pd \hat{\Phi}}{\pd \epsilon} \right\} \equiv \frac1{\eta^0}\frac{\pd}{\pd x} m(x,t) \\
    \dot{\Sigma} &= 2 \Laplace^w \Sigma \\
    \mu(0,t) = 0, \qquad \mu(1,t) = \lambda,& \qquad \Sigma(0,y,t) = \Sigma(1,y,t) = \Sigma(x,0,t) = \Sigma(x,1,t) = 0,
\end{align*}
where 
\begin{align*}
    \hat{\Phi}(x,t) &= \hat{\Phi}(\epsilon(x,t),\E(x,x,t))= \Phi\left(\frac{\epsilon(x,t)}{\sqrt{\E(x,x,t)}}\right)\\
    m(x,t) &= k_1^0(\epsilon + l_1^0)(1 - \hat{\Phi}) + k_2^0(\epsilon - l_2^0)\hat{\Phi} + (k_2^0 - k_1^0)\E(x,x,t)\frac{\pd \hat{\Phi}}{\pd \epsilon}\\
    \Laplace^w &= \frac1{2}\left( \frac{\pd}{\pd x} w(x,t) \frac{\pd}{\pd x} +\frac{\pd}{\pd y} w(y,t) \frac{\pd}{\pd y} \right) \\
    w(x,t) &= \frac{k_1^0}{\eta_0}(1 - \hat{\Phi}(x,t)) + \frac{k_2^0}{\eta_0}\hat{\Phi}(x,t)  - \frac{k_1^0l_1^0 + k_2^0l_2^0}{\eta_0}\frac{\pd \hat{\Phi}}{\pd \epsilon}(x,t).
\end{align*}
\par 
Having described the dynamics of the continuum limit, we now turn to describing the thermodynamics through the non-equilibrium free energy. From the previous section, we know that 
\begin{equation} \label{eqn:fe}
    (\hat{A}^\noneq)^{(N)}  = \sum_{i=1}^{N+1} U^{(N)}\left((D_1^{(N)}\mu^{(N)})_i,\sqrt{(D_1^{(N)}\Sigma^{(N)}(D_1^{(N)})^T)_{ii}}\right)  - \frac{N}{2\beta^{(N)}}(1 + \log(2\pi)) - \frac1{2\beta^{(N)}}\log(\det(\Sigma^{(N)})).
\end{equation}
where 
\begin{align*}
    U^{(N)}(\nu,\sigma)  &= \frac{k_1^{(N)}}{2}\left((\sigma^2 + (\nu + l^{(N)}_1)^2)\left(1 - \Phi(\nu/\sigma)\right) - \sigma(\nu + 2l^{(N)}_1)\phi(\nu/\sigma) \right) \\
    &\quad + \frac{k_2^{(N)}}{2}\left((\sigma^2 + (\nu-l_2^{(N)})^2)\Phi(\nu/\sigma) + \sigma(\nu - 2l_2^{(N)})\phi(\nu/\sigma)\right) + h_2^{(N)}\Phi(\nu/\sigma) .
\end{align*}
$l_1^{(N)}$, $l_2^{(N)}$, $D_1^{(N)}\mu^{(N)}_i$, $\sqrt{(D_1^{(N)}\Sigma^{(N)}(D_1^{(N)})^T)_{ii}}$, and $h_2^{(N)} = k_1^{(N)}(l_1^{(N)})^2/2 - k_2^{(N)}(l_2^{(N)})^2/2$ all scale like $1/N$. On the other hand, $k_1^{(N)}$ and $k_2^{(N)}$ both scale like $N$. Thus, $U^{(N)}\left((D_1^{(N)}\mu^{(N)})_i,\sqrt{(D_1^{(N)}\Sigma^{(N)}(D_1^{(N)})^T)_{ii}}\right)$ scales as $1/N$, and so 
\begin{equation*}
    \hat{E}(x,t) = \hat{E}(\mu(x,t),\Sigma(x,x,t)) \equiv \lim_{N\rightarrow\infty} \sum_{i=1}^{N+1} U^{(N)}\left((D_1\mu^{(N)})_i,\sqrt{(D_1\Sigma^{(N)}D_1^T)_{ii}}\right) = \int_0^1 U^0(\epsilon(x,t),\E(x,x,t))\diff x 
\end{equation*}
where 
\begin{align*}
    U^0(\epsilon,\E)&= \frac{k_1^{0}}{2}\left((\E + (\epsilon + l^{0}_1)^2)\left(1 - \hat{\Phi}\right) - \E(\epsilon + 2l^{0}_1)\frac{\pd \hat{\Phi}}{\pd \epsilon} \right) \\
    &\quad + \frac{k_2^{0}}{2}\left((\E + (\epsilon-l_2^{0})^2)\hat{\Phi} + \E(\epsilon - 2l_2^{0})\frac{\pd \hat{\Phi}}{\pd \epsilon} \right) + h_2^{0}\hat{\Phi}.
\end{align*}
Since 
\begin{align*}
    \frac{\pd U^0}{\pd \epsilon} &= k_1^0(\epsilon + l_1^0)\left(1 - \hat{\Phi}\right) + k_2^0(\epsilon - l_2^0)\hat{\Phi} + (k_2^0 - k_1^0)\E(x,x,t) \frac{\pd \hat{\Phi}}{\pd \epsilon} = m(x,t)\\ 
    \frac{\pd U^0}{\pd \E} &= \frac{k_1^0}{2}\left(1 - \hat{\Phi}\right) + \frac{k_2^0}{2}\hat{\Phi} - \frac1{2}(k_1^0 l_1^0 + k_2^0l_2^0) \frac{\pd \hat{\Phi}}{\pd \epsilon} = \frac{\eta^0}{2} w(x,t)
\end{align*}
we see that 
\begin{equation*}
    \frac{\delta \hat{A}^\noneq}{\delta \mu} = \frac{\delta \hat{E}}{\delta \mu} = -\frac{\pd}{\pd x} \frac{\pd U^0}{\pd \epsilon}(x,t) = -\eta \dot{\mu}
\end{equation*}
and we recover the gradient flow equation. 
To compute $\frac{\delta \hat{E}}{\delta \Sigma}$ we write
\begin{align*}
\hat{E}[\mu,\Sigma + \delta \Sigma] - \hat{E}[\mu,\Sigma] &= \int_0^1 U^0(\epsilon,\E + \delta \E) - U^0(\epsilon,\E)\diff x \\
    &= \int_0^1 \frac{\pd U^0}{\pd\E}\delta \E \diff x + \BigO(\delta \E^2) \\
    &= \int_0^1 \frac{\pd U^0}{\pd\E} \frac{\pd^2\delta \Sigma}{\pd x\pd y}(x,x,t) \diff x + \BigO(\delta \Sigma^2) \\ 
    &= \int_{[0,1]^2} \frac{\pd U^0}{\pd\E} \frac{\pd^2\delta \Sigma}{\pd x\pd y}(x,y,t)\delta(x -y) \diff x \diff y + \BigO(\delta \Sigma^2) \\
    &= \int_{[0,1]^2} \frac{\pd}{\pd x}\left( \frac{\pd U^0}{\pd \E} \frac{\pd}{\pd y}\delta(x-y)\right) \delta \Sigma(x,y,t)\diff x \diff y + \BigO(\delta \Sigma^2) \\
    &= \int_{[0,1]^2} \frac{\pd}{\pd x}\left( \frac{\eta^0}{2}w(x,t) \frac{\pd}{\pd y}\delta(x-y)\right) \delta \Sigma(x,y,t)\diff x \diff y + \BigO(\delta \Sigma^2) 
\end{align*}
Hence 
\begin{equation*}
    \frac{\delta \hat{E}}{\delta \Sigma}(x,y,t) = \frac{\pd}{\pd x}\left( \frac{\eta^0}{2}w(x,t)\frac{\pd}{\pd y}\delta(x-y)\right).
\end{equation*}
To recover the weighted Laplacian present in the dynamical equations for $\Sigma$, we integrate $\frac{\delta \hat{E}}{\delta \Sigma}$ against $\Sigma$, and test functions $u(x)$ and $v(y)$ (with compact support on $[0,1]$)
\begin{align*}
    \int\frac{\delta \hat{E}}{\delta \Sigma}(x,\xi,t)\Sigma(\xi,y,t)u(x)v(y)\diff \xi\diff x\diff y &= \int\frac{\pd}{\pd x}\left( \frac{\eta^0}{2}w(x,t)\frac{\pd}{\pd \xi}\delta(x-\xi)\right)\Sigma(\xi,y,t)u(x)v(y)\diff \xi \diff x \diff y \\ 
    &= -\int\left( \frac{\eta^0}{2}w(x)\frac{\pd}{\pd \xi}\delta(x-\xi)\right)\Sigma(\xi,y,t)u'(x)v(y)\diff \xi \diff x \diff y \\ 
    &= \int\left( \frac{\eta^0}{2}w(x,t)\right)\frac{\pd \Sigma(x,y,t)}{\pd x}u'(x)v(y) \diff x \diff y \\ 
    &= -\int \frac{\pd}{\pd x}\left( \frac{\eta^0}{2}w(x,t)\frac{\pd \Sigma(x,y,t)}{\pd x} \right)u(x)v(y) \diff x \diff y.
\end{align*}
So we see that 
\begin{equation*}
    \int\frac{\delta \hat{E}}{\delta \Sigma}(x,\xi,t)\Sigma(\xi,y,t)\diff \xi = - \frac{\pd}{\pd x}\left( \frac{\eta^0}{2}w(x,t)\frac{\pd \Sigma(x,y,t)}{\pd x} \right).
\end{equation*}
We use this to define the integration kernel
\begin{equation*}
    M(x,y,z,w,t) = \frac1{\eta^0}\left\{\Sigma(x,w,t)\delta(y-z) + \Sigma(w,y,t)\delta(x-z) + \Sigma(x,z,t)\delta(y-w) + \Sigma(y,z,t)\delta(x-w)\right\}
\end{equation*}
and by an analogous computation, we have 
\begin{equation*}
    \int M(x,y,z,w,t) \frac{\delta \hat{E}}{\delta \Sigma}(z,w,t) \diff z\diff w = -2 \Laplace^w \Sigma (x,y,t).
\end{equation*}
For the remaining portion of the free energy in Eqn. \ref{eqn:fe}, we assume that the functional derivative and continuum limit commute
\begin{equation*}
 \frac{\delta}{\delta \Sigma} \frac1{2\beta^{(N)}}\lim_{N\rightarrow \infty} \log(\det(\Sigma^{(N)})) = \lim_{N\rightarrow \infty} \frac{\pd}{\pd \Sigma_{ij}} \frac1{2\beta^{(N)}}\log(\det(\Sigma^{(N)})) = \lim_{N\rightarrow \infty} \frac1{2\beta^{(N)}}(\Sigma^{(N)})^{-1}_{ij}.
\end{equation*}
Finally, we also assume that application of the integral kernel $M$ to the limit on the right hand side of the previous equation is equal to the limit of the discretized integral operator acting on $\frac{(\Sigma^{(N)})^{-1}}{\beta^{(N)}}$
\begin{align*}
\int M(x,y,z,w,t) \left\{ \lim_{N\rightarrow\infty} \frac1{2\beta^{(N)}}(\Sigma^{(N)})^{-1} \right\}(z,w)\, \diff z\diff w &\\
    = \lim_{N\rightarrow\infty}  \frac1{N^2}\sum_{k,l=1}^N \frac1{N\eta^{(N)}}&\left(\Sigma_{il}^{(N)}\delta_{jk} + \Sigma_{lj}^{(N)}\delta_{ik} + \Sigma_{jk}^ {(N)}\delta_{il} + \Sigma_{ik}^{(N)}\delta_{jl} \right) \frac1{2\beta^{(N)}} (\Sigma^{(N)})^{-1}_{kl}.
\end{align*}
Computing this limit gives us
\begin{equation*}
\lim_{N\rightarrow\infty}  \frac1{N^2}\sum_{k=1}^N\frac{2\delta_{ik}\delta_{kj}}{\eta^0\beta^{(N)}} = \lim_{N\rightarrow \infty} \frac{2\delta_{ij}}{\eta^0\beta^0 N(N+1)} = 0.
\end{equation*}
Thus, under these assumptions, we see that $\frac{\delta\hat{E}}{\delta \Sigma} = \frac{\delta \hat{A}^\noneq}{\delta \Sigma}$ and so we recover the gradient flow equation for $\Sigma$
\begin{equation*}
\dot{\Sigma} = - \int M(x,y,z,w,t) \frac{\delta \hat{A}^\noneq}{\delta \Sigma}(z,w,t) \, \diff z\diff w.
\end{equation*}

\subsection{Location and velocity of the interface}
Now that we have the continuum evolution equations, we can use the internal variables to derive an equation for the location and velocity of the traveling wave front in the double-well mass-spring-chain system assuming the external protocol is such that a traveling front exists. We first note that the location of the front at a given time $t$ can be determined via the phase fraction as the point in the reference configuration for which the phase fraction at that given location and time is equal to $1/2$. Symbolically, if $I(t) \in [0,1]$ denotes the location of the phase front at time $t$, then we have 
\begin{equation*}
    \hat{\Phi}\left(\epsilon(I(t),t),\E(I(t),I(t),t)\right) = \frac1{2}.
\end{equation*}
Since $\hat{\Phi}$ is given by the cumulative distribution function of a standard Gaussian, and $\E(x,x,t),\, \sqrt{\E(x,x,t)} > 0$ for all time assuming it is initially positive, the previous equation is true if and only if $\epsilon(I(t),t) = 0$. Now, we assume that $\epsilon(x,t)$ is invertable for $x \in[0,1]$ for all times. This ensures that there is only one phase front propagating through the system. It's equally valid to assume that $\mu(x,t)$ is strictly convex (or concave), which we observe to be the case whenever there is a single propagating phase front (if there were multiple phase fronts, the following analysis could be preformed on each convex component of $\mu$ in order to determine each front location). Now, by differentiating the equation $\epsilon(I(t),t) = 0$ in time and rewriting and arranging terms, we find that
\begin{equation*}
    \dot{I}(t) = -\frac{\dot{\epsilon}(I(t),t)}{\frac{\pd \epsilon}{\pd x}(I(t),t)} .
\end{equation*}
Next, we use the definition of $\epsilon = \frac{\pd \mu}{\pd x}$, and the gradient flow equation for $\dot{\mu}$ to get 
\begin{align*}
    \dot{I}(t) &= -\frac1{\frac{\pd^2 \mu}{\pd x^2}(I(t),t)} \frac{\pd^2 }{\pd t\pd x} \mu(I(t),t) \\ 
    &= \frac1{\eta\frac{\pd^2 \mu}{\pd x^2}(x,t)} \frac{\pd }{\pd x} \frac{\delta \hat{A}^\noneq}{\delta \mu}\Big|_{x = I(t)} \\
    &= -\frac1{\eta\frac{\pd^2 \mu}{\pd x^2}(x,t)} \frac{\pd^2 }{\pd x^2} \frac{\delta \hat{A}^\noneq}{\delta \epsilon}\Big|_{x = I(t)}. \\
\end{align*}
Each of these equations is an equally valid ordinary differential equation for the phase front location in terms of the internal variables and the non-equilibrium free energy. The final equation, however, 
\begin{equation*}
   \dot{I}(t) = -\frac1{\eta\frac{\pd^2 \mu}{\pd x^2}(x,t)} \frac{\pd^2 }{\pd x^2} \frac{\delta \hat{A}^\noneq}{\delta \epsilon}\Big|_{x = I(t)}
\end{equation*}
reveals that the phase front is given by the ratio of the curvature of the thermodynamic affinity conjugate to the strain $\mathcal{A}_\epsilon \equiv - \frac{\delta \hat{A}^{\noneq}}{\delta \epsilon}$ and the curvature of $\mu$ at the location of the phase front.

\subsection{Relevant Gaussian integrals} \label{sec:gaussInt}
In this section, we compute the relevant Gaussian integrals necessary for the double-well mass-spring-chain example. These quantities include various expectations of the interaction energy $u(z)$ and its derivative $u'(z)$ for separations $z$ which obey a Gaussian distribution. 
Suppose $\Xi$ and $\Psi$ are jointly Gaussian with means $\mu_\xi$ and $\mu_\psi$ and covariance $\Sigma = \mat{ \sigma_\xi^2 & \rho \sigma_\xi \sigma_\psi \\ \rho \sigma_\xi \sigma_\psi & \sigma_\psi^2 \\ }$. 
Let $\Phi(z)$ and $\phi(z)$ be the cumulative distribution function (cdf) and probability density function (pdf) of a standard normal random variable. 
Also recall the definition
\begin{equation*}
    u(z) = \begin{cases} \frac{k_1}{2}(z + l_1)^2 & z \leq 0 \\ \frac{k_2}{2}(z - l_2)^2 + h_2& z > 0 \\ \end{cases} 
\end{equation*}
for the interaction energy and its derivative
\begin{equation*}
    u'(z) = \begin{cases} k_1(z + l_1) & z \leq 0 \\ k_2(z - l_2) & z > 0 \\ \end{cases}.
\end{equation*}
Our goal is to compute $\avg{u'(\Xi)}$, $\avg{u'(\Xi)\Xi}$, $\avg{u'(\Xi)\Psi}$, and $\avg{u(\Xi)}$ where, in this section, $\avg{\cdots}$ denotes averaging with respect to the joint distribution of $\Xi$ and $\Psi$. 
First, we can compute $\avg{u'(\Xi)}$ and $\avg{u'(\Xi)\Xi}$ through straightforward integration to get 
\begin{align*}
    \avg{u'(\Xi)} &= \int_{-\infty}^0 k_1(\xi + l_1)\frac{\phi\left(\frac{\xi - \mu_\xi}{\sigma_\xi}\right)}{\sigma_\xi}\,\diff \xi + \int_0^\infty k_2(\xi - l_2)\frac{\phi\left(\frac{\xi - \mu_\xi}{\sigma_\xi}\right)}{\sigma_\xi}\,\diff \xi \\
    &= \int_{-\infty}^{-\mu_\xi/\sigma_\xi} k_1(\sigma_\xi \psi +\mu_\xi + l_1)\phi(\psi)\,\diff \psi + \int_{-\mu_\xi/\sigma_\xi}^{\infty} k_2(\sigma_\xi \psi +\mu_\xi - l_2)\phi(\psi)\,\diff \psi \\
    &= k_1\left(-\sigma_\xi \phi(-\mu_\xi/\sigma_\xi) + (\mu_\xi + l_1)\Phi(-\mu_\xi/\sigma_\xi) \right) + k_2\left(\sigma_\xi\phi(\mu_\xi/\sigma_\xi) + (\mu_\xi - l_2)\Phi(\mu_\xi/\sigma_\xi) \right),
\end{align*}
and
\begin{align*}
    \avg{u'(\Xi)\Xi} &=\int_{-\infty}^0 \xi\ k_1(\xi + l_1)\frac{\phi\left(\frac{\xi - \mu_\xi}{\sigma_\xi}\right)}{\sigma_\xi}\,\diff \xi + \int_0^\infty \xi\ k_2(\xi - l_2)\frac{\phi\left(\frac{\xi - \mu_\xi}{\sigma_\xi}\right)}{\sigma_\xi}\,\diff \xi \\
    &= \int_{-\infty}^{-\mu_\xi/\sigma_\xi}(\sigma_\xi \psi+\mu_\xi) k_1(\sigma_\xi \psi +\mu_\xi + l_1)\phi(\psi)\,\diff \psi + \int_{-\mu_\xi/\sigma_\xi}^{\infty} (\sigma_\xi \psi+\mu_\xi) k_2(\sigma_\xi \psi +\mu_\xi - l_2)\phi(\psi)\,\diff \psi \\
    &= k_1\left\{(\sigma_\xi^2+\mu_\xi(\mu_\xi+l_1))\Phi(-\mu_\xi/\sigma_\xi)- \sigma_\xi(\mu_\xi + l_1)\phi(-\mu_\xi/\sigma_\xi) \right\} + k_2\left\{(\sigma_\xi^2 + \mu_\xi(\mu_\xi-l_2)) \Phi(\mu_\xi/\sigma_\xi) + \sigma_\xi(\mu_\xi-l_2)\phi(\mu_\xi/\sigma_\xi)\right\}
\end{align*}
as $\Phi'(\xi) = \phi(\xi)$, $\phi'(\xi) = -\xi\phi(\xi)$, $\Phi(\xi) = 1 - \Phi(-\xi)$, and $\phi(\xi) = \phi(-\xi)$.
\par
Next, we'd like to compute $\avg{u'(\Xi)\Psi}$. To do so, we make use of the fact that if $Z_1$ and $Z_2$ are independent standard Gaussians (mean zero, variance one), then $\Xi$ and $\Psi$ are jointly equal in distribution to $(\Xi,\Psi) =_D (\sigma_\xi Z_1 + \mu_\xi,\ \rho \sigma_\psi Z_1 + \sqrt{(1 - \rho^2)}\sigma_\psi\ Z_2 + \mu_\psi)$. Thus, if we define $\Psi^\perp$ to be a Gaussian with mean zero, variance $\sigma_\psi^2(1 - \rho^2)$ and independent of $\Xi$, we can also write $(\Xi,\Psi) =_D (\Xi,\frac{\rho\sigma_\psi}{\sigma_\xi}(\Xi - \mu_\xi) + \mu_\psi + \Psi^\perp)$. And so, we use the linearity of expectations to compute 
\begin{align*}
    \avg{u'(\Xi)\Psi} &= \avg{u'(\Xi)\left(\frac{\rho \sigma_\psi}{\sigma_\xi}(\Xi - \mu_\xi) + \mu_\psi + \Psi^\perp\right)}\\
    &=\frac{\rho\sigma_\psi}{\sigma_\xi}\avg{u'(\Xi)\Xi} + \left(\mu_\psi - \frac{\rho\sigma_\psi}{\sigma_\xi}\mu_\xi\right)\avg{u'(\Xi)} + \avg{u'(\Xi)\ \Psi^\perp} \\
    &= \frac{\rho\sigma_\psi}{\sigma_\xi}\Big[k_1\left\{(\sigma_\xi^2+\mu_\xi(\mu_\xi+l_1))\Phi(-\mu_\xi/\sigma_\xi)- \sigma_\xi(\mu_\xi + l_1)\phi(-\mu_\xi/\sigma_\xi) \right\} \\
    &\quad + k_2\left\{(\sigma_\xi^2 + \mu_\xi(\mu_\xi-l_2)) \Phi(\mu_\xi/\sigma_\xi) + \sigma_\xi(\mu_\xi-l_2)\phi(\mu_\xi/\sigma_\xi)\right\}\Big]\\
    &\quad + \left(\mu_\psi - \frac{\rho\sigma_\psi}{\sigma_\xi}\mu_\xi\right)\left(k_1\left(-\sigma_\xi \phi(-\mu_\xi/\sigma_\xi) + (\mu_\xi + l_1)\Phi(-\mu_\xi/\sigma_\xi) \right) + k_2\left(\sigma_\xi\phi(\mu_\xi/\sigma_\xi) + (\mu_\xi - l_2) \Phi(\mu_\xi/\sigma_\xi) \right)\right) \\
    &=k_1 \bigg\{ (\rho\sigma_\xi\sigma_\psi + \mu_\psi(\mu_\xi + l_1))\Phi(-\mu_\xi/\sigma_\xi) - (\mu_\psi\sigma_\xi+\rho\sigma_\psi l_1  )\phi(-\mu_\xi/\sigma_\xi)\bigg\} \\
    &\quad + k_2\bigg\{ (\rho\sigma_\xi\sigma_\psi + \mu_\psi(\mu_\xi - l_2))\Phi(\mu_\xi/\sigma_\xi) + (\mu_\psi\sigma_\xi -\rho\sigma_\psi l_2)\phi(\mu_\xi/\sigma_\xi)\bigg\}
\end{align*}
In the special case of $\mu_\psi = 0$, which we will need, we have 
\begin{equation*}
    \avg{u'(\Xi)\Psi} = \Sigma_{\xi\psi}\left(k_1\left(1 - \Phi\left(\frac{\mu_\xi}{\sqrt{\Sigma_{\xi\xi}}}\right)\right) + k_2\Phi\left(\frac{\mu_\xi}{\sqrt{\Sigma_{\xi\xi}}}\right)-(k_1l_1 + k_2l_2)\frac{\phi\left(\frac{\mu_\xi}{\sqrt{\Sigma_{\xi\xi}}}\right)}{\sqrt{\Sigma_{\xi\xi}}} \right).
\end{equation*}
In the double-well mass-spring-chain example, we assume the $N$ internal mass positions, $x$, are modeled as a multivariate normal with mean $\mu$ and covariance $\Sigma$. We shall need to take the specific case of $\Xi = (D_1x)_i$ and $\Psi = (x-\mu)_j$. In this case, we have $\mu_\xi = (D_1\mu)_i$, $\mu_\psi = 0$, $\Sigma_{\xi\xi} = (D_1\Sigma D_1^T)_{ii} = \tau_i^2$, and $\Sigma_{\xi\psi} = D_1\Sigma$ so that 
\begin{align*}
    \avg{u'((D_1x)_i)(x-\mu)_j}_{\hat{p}} &= \avg{u'(\Xi)\Psi}_{\Xi,\Psi} = (D_1\Sigma)_{ij}\left(k_1\left(1 - \Phi\left(\frac{(D_1\mu)_i}{\tau_i}\right)\right) + k_2\Phi\left(\frac{(D_1\mu)_i}{\tau_i}\right) - (k_1l_1 + k_2l_2)\frac{\phi\left(\frac{(D_1\mu)_i}{\tau_i}\right)}{\tau_i} \right) \\
    &\equiv \eta (D_1\Sigma)_{ij} w_i
\end{align*}
for $i,j = 1,\ldots,N$. When $i = N+1$, we simply get $\avg{u'((D_1x)_i)(x-\mu)_j}_{\hat{p}} = -\eta \Sigma_{Nj}\, w_{N+1}$ where $w_{N+1}$ is defined analogously to $w_i$. 
\par
Finally, by an analogous computation to those above, one can find that (we drop the $x$ subscript since it is no longer needed)
\begin{align*}
    U(\nu,\sigma) \equiv \avg{u(\Xi)}_{\Xi\sim N(\nu,\sigma^2)} &= \frac{k_1}{2}\left((\sigma^2 + (\nu + l_1)^2)\Phi(-\nu/\sigma) - \sigma(\nu + 2l_1)\phi(-\nu/\sigma) \right) \nonumber \\
    &\quad + \frac{k_2}{2}\left((\sigma^2 + (\nu-l_2)^2)\Phi(\nu/\sigma) + \sigma(\nu - 2l_2)\phi(\nu/\sigma) \right) + h_2 \Phi(\nu/\sigma)
\end{align*}
It will be useful to have at hand 
\begin{align*}
    \frac{\pd}{\pd \nu}U(\nu,\sigma) &= \frac{k_1}{2}\left(2(\nu+l_1)\Phi(-\nu/\sigma) - \frac1{\sigma}(l_1^2+2\sigma^2)\phi(-\nu/\sigma) \right) \nonumber \\
        &\quad + \frac{k_2}{2}\left(2(\nu-l_2)\Phi(\nu/\sigma) + \frac1{\sigma}(l_2^2 + 2\sigma^2)\phi(\nu/\sigma) \right) + \frac{h_2}{\sigma}\phi(\nu/\sigma) \nonumber \\
        &= k_1(\nu + l_1)\Phi(-\nu/\sigma) + k_2(\nu - l_2)\Phi(\nu/\sigma) + \sigma(k_2 - k_1)\phi(\nu/\sigma) \\
        \nonumber\\
    \frac{\pd}{\pd \sigma}U(\nu,\sigma) &= \frac{k_1}{2}\left(2\sigma\Phi(-\nu/\sigma) + \frac{\nu l_1^2}{\sigma^2}\phi(-\nu/\sigma) - 2l_1\phi(-\nu/\sigma) \right) \nonumber \\
    &\quad + \frac{k_2}{2}\left(2\sigma\Phi(\nu/\sigma) - \frac{\nu l_2^2}{\sigma^2}\phi(\nu/\sigma) - 2l_2\phi(\nu/\sigma) \right) - \frac{\nu h_2}{\sigma^2}\phi(\nu/\sigma) \nonumber \\
    &= k_1 \sigma \Phi(-\nu/\sigma) + k_2 \sigma \Phi(\nu/\sigma) - (k_1l_1 + k_2l_2)\phi(\nu/\sigma)
\end{align*}
where cancellation occurs because we have assumed $u(z)$ is continuous and hence $\frac{k_2}{2}l_2^2 + h_2 - \frac{k_1}{2}l_1^2 = 0$.

\subsection{Langevin simulations}
In order to check our analytical results, we have compared our predictions to Langevin simulations of the single particle and double-well mass-spring-chain system. Throughout, we have approximated solutions to the governing stochastic differential equation using the standard first order Euler-Maruyama scheme. That is, if the true solution $x(t)$ obeys 
\begin{equation*}
    \,\diff x(t) = \mu(x(t),t)\,\diff t + \sigma \,\diff b(t)
\end{equation*}
for a known drift $\mu(x,t)$ and homogeneous, stationary diffusion $\sigma$, we approximate the finite differences as 
\begin{equation*}
    x(t_{i+1}) = x(t_i) + \mu(x(t_i),t_i)\Delta t + \sigma \sqrt{\Delta t}\, \xi_i
\end{equation*}
with $\Delta t = 10^{-5}$, $t_i = i \Delta t$, and $\xi_i$ are independent, standard normals. Since the diffusion is homogeneous, the method is $\BigO(\Delta t)$ accurate (with probability one). 
We should also note that all ordinary differential equations for the internal variables $\dot{\alpha}(t) = f(\alpha(t),t)$ are solved using a standard fourth order Runge-Kutta method with $\Delta t = 10^{-2}$.
\par 
Most of the marcoscopic quantities presented in the main text can be approximated from the Langevin simulations using empirical averages with respect to the samples. Mathematically, if we simulate sample trajectories $\{x^{(i)}(t^{(j)}) \mid i = 1,\ldots,N_\text{data} \text{ and } j =1,\ldots,N_\text{times}\}$, we can approximate 
\begin{equation*}
    \avg{g(x,t^{(j)})}_p = \int g(x,t^{(j)})p(x,t^{(j)} \mid \lambda) \,\diff x \approx \frac1{N_\text{data}}\sum_{i=1}^{N_\text{data}} g(x^{(i)}(t^{(j)}),t^{(j)}) \qquad j = 1,\ldots,N_\text{times}.
\end{equation*}

However, the rate of total entropy production poses a challenge in the double-well mass-spring-chain example. We make use of the splitting of the total entropy production into the entropy of the medium, and configurational entropy, $\Delta s^\text{tot} = \Delta s^\text{m} + \Delta s$. Furthermore, we use of the identities $\Delta s^\text{m} = \frac{\Delta q}{T} = \frac{\Delta w - \Delta e}{T}$, in order to calculate the change in entropy of the medium. The internal energy $e$, is a state function and so its average can be computed at each time step, and the difference between time steps used to compute $\Delta e$. The work done can also be computed using the discretization of its stochastic differential equation 
\begin{equation*}
\diff w(t) = \frac{\pd e}{\pd \lambda}\dot{\lambda}\,\diff t \quad \rightarrow \quad \Delta w(t) \approx \frac{\pd e}{\pd \lambda}(x(t),\lambda(t))\dot{\lambda}(t)\Delta t.
\end{equation*}
Both $\Delta e$ and $\Delta w$ are then averaged over samples to obtain $\Delta S^\text{m} = \frac{\Delta W - \Delta E}{T}$. 
\par 
For the entropy, $s(x,t) = -k_B \log(p(x,t))$, we approximate $p(x,t)$ using Gaussian kernel density estimation (GKDE) \cite{kim2012robust}. At each time step, $t^{(j)}$, we split the samples $\{x^ {(i)} = x^{(i)}(t^{(j)})\}_{i=1}^{N_\text{data}}$ into a maximum likelihood set $X_\text{MLE} = \{x^ {(i)}\}_{i=1}^{N_\text{MLE}}$, a training set $X_\text{train} = \{x^{(i)}\}_{i=N_\text{MLE} + 1}^{N_\text{MLE} + N_\text{train}}$, and an evaluation set $X_\text{eval} = \{x^{(i)}\}_{i=N_\text{MLE} + N_\text{train} + 1}^{N_\text{data}}$. We use $N_\text{MLE} = .1\ N_\text{data}$, $N_\text{train} = N_\text{eval} = .45\ N_\text{data}$. GKDE approximates the true density of the samples, $p(x,t^{(j)}\mid\lambda)$, using the one parameter family given by
\begin{equation*}
    p(y \mid X_\text{train}, h) = \frac1{N_\text{train}}\sum_{x \in X_\text{train}} \frac{\phi\left(\frac{y - x}{h}\right)}{h^d}
\end{equation*}
where $\phi(x) = (2\pi)^{-d/2}\exp(-\norm{x}^2/2)$ is the density of a standard multivariate normal in $\bb{R}^d$. In words, this family of densities places a isotropic Gaussian density at each sample point $x \in X_\text{train}$ whose width is given by the parameter $h >0$. 
We then use the $X_\text{MLE}$ samples to preform maximum likelihood estimation on the optimal $h$. We take 
\begin{equation*}
    h_\text{MLE} = \text{argmax}_{h >0} \sum_{x \in X_\text{MLE}} \log(p(x\mid X_\text{train},h)).
\end{equation*}
Finally, we approximate $S(t^{(j)})$ by averaging over the evaluation set
\begin{equation*}
    S(t_j) = -k_B\frac1{N_\text{eval}}\sum_{x \in X_\text{eval}} \log(p(x\mid X_\text{train},h_\text{MLE})), 
\end{equation*}
and use differences in time to estimate $\Delta S$. 

\subsection{Front velocity and rate of entropy production due to the phase front}
\begin{figure}[h!]
    \centering
    \includegraphics{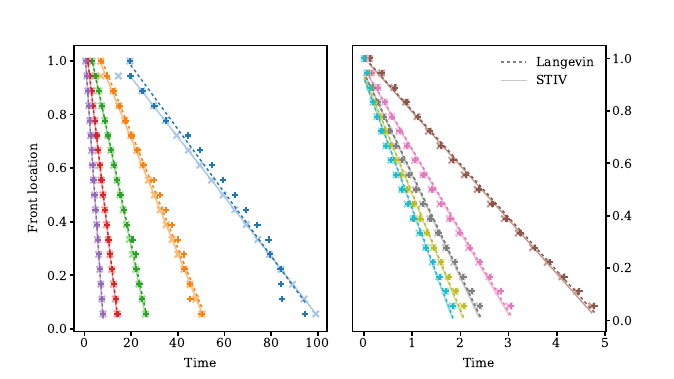}
    \caption{Phase front location in the reference configuration as a function of time. Light colored solid lines and x markers denote results from the STIV framework. Darker colored dashed lines and $+$ markers denote those from Langevin simulations (averaged over 100,000 simulations). Scattered points denote the time at which the phase front crosses a given spring. Least-squares best lines are fit to each data set the absolute value of the slope is taken as the phase front speed.}
    \label{fig:PhaseVel}
\end{figure}

Here, we give the details for computing the phase front velocity and rate of entropy production due to the phase front in the double-well mass-spring-chain system. Recall that we have defined the spring interaction potential so that one minima falls on the left of the origin and one on the right with the origin as the cross over point. Mathematically, the interaction is given by
\begin{equation*}
    u(z) = \begin{cases} \frac{k_1}{2}(z + l_1)^2 & z \leq 0 \\ 
    \frac{k_2}{2}(z - l_2)^2 + h_2 & z > 0, \\ 
    \end{cases}
\end{equation*}
where $h_2 = (k_1l_1^2 - k_2l_2^2)/2$ is fixed so that $u(z)$ is continuous. In order to induce a travelling phase front, we initialize the external protocol to $\lambda(0) = -1.75(N+1)l_1$ so that the minimum energy configuration has all springs falling far to the left of the negative minima. $\lambda(t)$ then increases with constant strain rate (values between $1/16$ and $10$ were used), and one by one the springs cross the barrier at the origin to fall within the positive minima, starting with the spring attached to the external protocol. At spring $i$, the predicted phase fraction from the STIV formalism is given by $\hat{\Phi}_i(t) = \Phi\left(\frac{(D_1\mu)_i(t)}{\sqrt{(D_1\Sigma(t) D_1^T)_{ii}}}\right)$ whereas the observed phase fraction from Langevin simulations can be estimated empirically as 
\begin{equation*}
    \Phi^L_i(t) = \frac1{N_\text{data}}\sum_{j=1}^{N_\text{data}} \theta\big(x_{i+1}^{(j)}(t) - x_i^{(j)}(t)\big),
\end{equation*}
where $N_\text{data}$ is the number of simulations (100,000 in our case) and $x^{(j)}_i(t)$ is the location of the $i$th mass in the $j$th simulation at time $t$, and $\theta(x)$ is the Heaviside function. The phase front is considered to be located at spring $i$ at the time when $\hat{\Phi}_i(t) = .5$ for STIV and $\Phi^L_i(t) = .5$ for the Langevin simulations. In the reference configuration, spring $i$ has position $\frac{i}{N+1}$. In Fig. \ref{fig:PhaseVel}, the times at which the phase front reaches each spring is plotted against each spring's position in the reference configuration for ten strain rates. In both the Langevin and STIV data, we see that the phase front has a roughly constant velocity. We take the absolute value of the slope of a least squares linear fit through each data set as the phase front speed at the given strain rate.
\par
In order to estimate the rate of entropy production due to the phase front, we take the total rate of entropy production and subtract of the contribution due to viscous drag resulting from movement of the mean mass location. For the true rate of entropy production (estimated using Langevin simulations), this gives
\[ T\frac{d}{dt}S^\text{phase} = T\frac{d}{dt}S^\text{tot} - \eta \norm{\frac{d}{dt} X}^2 \]
where $X_i = \avg{x}_i$. 
For STIV, we have  
\[ T\frac{d}{dt}\hat{S}^\text{phase} = T\frac{d}{dt}\hat{S}^\text{tot} - \eta \norm{\dot{\mu}}^2 = - \frac{\pd \hat{A}^\noneq}{\pd \Sigma^{-1}} : \dot{\Sigma}^{-1}. \]

\subsection{Movie S1}\label{Movie S1}
A comparison of the approximate STIV density $\hat{p}(x, \mu, \Sigma^{-1})$ (depicted as grey scale contour lines) and a histogram of simulated Langevin trajectories for a mass-spring-chain system with double well interactions with two degrees of freedom. The two axes show the lengths of the first and second spring. The simulation parameters are $\beta = 1$, $\eta = 1$, $k_1 = 1$, $l_1 = 1$, $k_2 = 4$, and $l_2 = 1/2$. The height of the barriers are $b_1 = k_1 l_1^2/2 = 1/2$ and $b_2 = k_2l_2^2/2 = 1/2$. Note $b_1,b_2 < 1/\beta$ in this case. Although the underlying histogram is multimodal, the modes fall within a single ``basin'' which is well approximated by the STIV density.
\begin{center}
    \includemedia[width = .6\textwidth,height = .3375\textwidth]{Movie S1}{img/MovieS1.swf}
\end{center}

\subsection{Movie S2}\label{Movie S2}
A comparison of the approximate STIV density $\hat{p}(x, \mu, \Sigma^{-1})$ (depicted as grey scale contour lines) and a histogram of simulated Langevin trajectories for a mass-spring-chain system with double well interactions with two degrees of freedom. The two axes show the lengths of the first and second spring. The simulation parameters are $\beta = 1$, $\eta = 1$, $k_1 = 1$, $l_1 = 1$, $k_2 = 8$, and $l_2 = 2$. The height of the barriers are $b_1 = k_1 l_1^2/2 = 1/2$ and $b_2 = k_2l_2^2/2 = 16$. Now $b_2 >> 1/\beta$. We see the simulation points of the histogram transition between different peaks, as well as the clear separation of modes. In this case, the STIV approximation appears to capture the behavior of only the dominant mode.
\begin{center}
    \includemedia[width = .6\textwidth,height = .3375\textwidth]{Movie S2}{img/MovieS2.swf}
\end{center}



\begin{thebibliography}{10}

    \bibitem{casella2021statistical}
    George Casella and Roger~L Berger.
    \newblock {\em Statistical inference}.
    \newblock Cengage Learning, 2021.
    
    \bibitem{coleman1964}
    B.D. Coleman.
    \newblock Thermodynamics of materials with memory.
    \newblock {\em Arch. Rat. Mech. Anal.}, 17:1--46, 1964.
    
    \bibitem{connolly2009geodynamic}
    JAD Connolly.
    \newblock The geodynamic equation of state: what and how.
    \newblock {\em Geochemistry, Geophysics, Geosystems}, 10(10), 2009.
    
    \bibitem{crooks1999}
    Gavin~E Crooks.
    \newblock Entropy production fluctuation theorem and the nonequilibrium work
      relation for free energy differences.
    \newblock {\em Physical Review E}, 60(3):2721, 1999.
    
    \bibitem{doi2011onsager}
    Masao Doi.
    \newblock Onsager’s variational principle in soft matter.
    \newblock {\em Journal of Physics: Condensed Matter}, 23(28):284118, 2011.
    
    \bibitem{dunatunga2015continuum}
    Sachith Dunatunga and Ken Kamrin.
    \newblock Continuum modelling and simulation of granular flows through their
      many phases.
    \newblock {\em Journal of Fluid Mechanics}, 779:483--513, 2015.
    
    \bibitem{eyink1996}
    Gregory~L Eyink.
    \newblock Action principle in nonequilibrium statistical dynamics.
    \newblock {\em Physical Review E}, 54(4):3419, 1996.
    
    \bibitem{feng2006large}
    Jin Feng and Thomas~G Kurtz.
    \newblock {\em Large deviations for stochastic processes}.
    \newblock Number 131. American Mathematical Soc., 2006.
    
    \bibitem{gompper2020}
    Gerhard Gompper, Roland~G Winkler, Thomas Speck, Alexandre Solon, Cesare
      Nardini, Fernando Peruani, Hartmut L{\"o}wen, Ramin Golestanian, U~Benjamin
      Kaupp, Luis Alvarez, et~al.
    \newblock The 2020 motile active matter roadmap.
    \newblock {\em Journal of Physics: Condensed Matter}, 32(19):193001, 2020.
    
    \bibitem{gore2006}
    Jeff Gore, Zev Bryant, Marcelo N{\"o}llmann, Mai~U Le, Nicholas~R Cozzarelli,
      and Carlos Bustamante.
    \newblock Dna overwinds when stretched.
    \newblock {\em Nature}, 442(7104):836--839, 2006.
    
    \bibitem{grmela1997dynamics}
    Miroslav Grmela and Hans~Christian {\"O}ttinger.
    \newblock Dynamics and thermodynamics of complex fluids. i. development of a
      general formalism.
    \newblock {\em Physical Review E}, 56(6):6620, 1997.
    
    \bibitem{gupta2021nonequilibrium}
    Prateek Gupta, Michael Ortiz, and Dennis~M Kochmann.
    \newblock Nonequilibrium thermomechanics of gaussian phase packet crystals:
      Application to the quasistatic quasicontinuum method.
    \newblock {\em Journal of the Mechanics and Physics of Solids}, 153:104495,
      2021.
    
    \bibitem{gurtin2010mechanics}
    Morton~E Gurtin, Eliot Fried, and Lallit Anand.
    \newblock {\em The mechanics and thermodynamics of continua}.
    \newblock Cambridge University Press, 2010.
    
    \bibitem{heller1975time}
    Eric~J Heller.
    \newblock Time-dependent approach to semiclassical dynamics.
    \newblock {\em The Journal of Chemical Physics}, 62(4):1544--1555, 1975.
    
    \bibitem{hemminger2007directing}
    John Hemminger, Graham Fleming, and M~Ratner.
    \newblock Directing matter and energy: Five challenges for science and the
      imagination.
    \newblock Technical report, DOESC (USDOE Office of Science (SC)), 2007.
    
    \bibitem{horowitz2020}
    Jordan~M Horowitz and Todd~R Gingrich.
    \newblock Thermodynamic uncertainty relations constrain non-equilibrium
      fluctuations.
    \newblock {\em Nature Physics}, 16(1):15--20, 2020.
    
    \bibitem{horstemeyer2010}
    Mark~F Horstemeyer and Douglas~J Bammann.
    \newblock Historical review of internal state variable theory for inelasticity.
    \newblock {\em International Journal of Plasticity}, 26(9):1310--1334, 2010.
    
    \bibitem{jaeger2010far}
    Heinrich~M Jaeger and Andrea~J Liu.
    \newblock Far-from-equilibrium physics: An overview.
    \newblock {\em arXiv preprint arXiv:1009.4874}, 2010.
    
    \bibitem{jarzynski1997}
    Christopher Jarzynski.
    \newblock Nonequilibrium equality for free energy differences.
    \newblock {\em Physical Review Letters}, 78(14):2690, 1997.
    
    \bibitem{jou1996}
    David Jou, Jos{\'e} Casas-V{\'a}zquez, and Georgy Lebon.
    \newblock Extended irreversible thermodynamics.
    \newblock In {\em Extended Irreversible Thermodynamics}, pages 41--74.
      Springer, 1996.
    
    \bibitem{kim2012robust}
    JooSeuk Kim and Clayton~D Scott.
    \newblock Robust kernel density estimation.
    \newblock {\em The Journal of Machine Learning Research}, 13(1):2529--2565,
      2012.
    
    \bibitem{kreplak2004new}
    L~Kreplak, J~Doucet, P~Dumas, and F~Briki.
    \newblock New aspects of the $\alpha$-helix to $\beta$-sheet transition in
      stretched hard $\alpha$-keratin fibers.
    \newblock {\em Biophysical Journal}, 87(1):640--647, 2004.
    
    \bibitem{kulkarni2008variational}
    Yashashree Kulkarni, Jaroslaw Knap, and Michael Ortiz.
    \newblock A variational approach to coarse graining of equilibrium and
      non-equilibrium atomistic description at finite temperature.
    \newblock {\em Journal of the Mechanics and Physics of Solids},
      56(4):1417--1449, 2008.
    
    \bibitem{complexDynamicsBAA}
    Army~Research Lab.
    \newblock Complex dynamics and systems.
    \newblock {\em ARL Broad Agency Announcement}, 2020.
    
    \bibitem{lebon2008understanding}
    Georgy Lebon, David Jou, and Jos{\'e} Casas-V{\'a}zquez.
    \newblock {\em Understanding non-equilibrium thermodynamics}, volume 295.
    \newblock Springer, 2008.
    
    \bibitem{li2011diffusive}
    Ju~Li, Sanket Sarkar, William~T Cox, Thomas~J Lenosky, Erik Bitzek, and Yunzhi
      Wang.
    \newblock Diffusive molecular dynamics and its application to nanoindentation
      and sintering.
    \newblock {\em Physical Review B}, 84(5):054103, 2011.
    
    \bibitem{li2019harnessing}
    Xiaoguai Li, Nicolas Dirr, Peter Embacher, Johannes Zimmer, and Celia Reina.
    \newblock Harnessing fluctuations to discover dissipative evolution equations.
    \newblock {\em Journal of the Mechanics and Physics of Solids}, 131:240--251,
      2019.
    
    \bibitem{maugin1994}
    Gerard~A Maugin and Wolfgang Muschik.
    \newblock Thermodynamics with internal variables. part i. general concepts.
    \newblock {\em Journal of Non-Equilibrium Thermodynamics}, 19:217--249, 1994.
    
    \bibitem{maugin1994thermodynamics2}
    G{\'e}rard~A Maugin and Wolfgang Muschik.
    \newblock Thermodynamics with internal variables. part ii. applications.
    \newblock {\em Journal of Non-Equilibrium Thermodynamics}, 19:250--289, 1994.
    
    \bibitem{mielke2011formulation}
    Alexander Mielke.
    \newblock Formulation of thermoelastic dissipative material behavior using
      generic.
    \newblock {\em Continuum Mechanics and Thermodynamics}, 23(3):233--256, 2011.
    
    \bibitem{mielke2016generalization}
    Alexander Mielke, DR~Michiel Renger, and Mark~A Peletier.
    \newblock A generalization of onsager’s reciprocity relations to gradient
      flows with nonlinear mobility.
    \newblock {\em Journal of Non-Equilibrium Thermodynamics}, 41(2):141--149,
      2016.
    
    \bibitem{montefusco2021framework}
    Alberto Montefusco, Mark~A Peletier, and Hans~Christian {\"O}ttinger.
    \newblock A framework of nonequilibrium statistical mechanics. ii.
      coarse-graining.
    \newblock {\em Journal of Non-Equilibrium Thermodynamics}, 46(1):15--33, 2021.
    
    \bibitem{nemat2004plasticity}
    Siavouche Nemat-Nasser.
    \newblock {\em Plasticity: a treatise on finite deformation of heterogeneous
      inelastic materials}.
    \newblock Cambridge University Press, 2004.
    
    \bibitem{onsager1931reciprocal}
    Lars Onsager.
    \newblock Reciprocal relations in irreversible processes. i.
    \newblock {\em Physical Review}, 37(4):405, 1931.
    
    \bibitem{ortiz1999variational}
    Michael Ortiz and Laurent Stainier.
    \newblock The variational formulation of viscoplastic constitutive updates.
    \newblock {\em Computer methods in applied mechanics and engineering},
      171(3-4):419--444, 1999.
    
    \bibitem{ottinger1998}
    Hans~Christian {\"O}ttinger.
    \newblock General projection operator formalism for the dynamics and
      thermodynamics of complex fluids.
    \newblock {\em Physical Review E}, 57(2):1416, 1998.
    
    \bibitem{ottinger2005}
    Hans~Christian {\"O}ttinger.
    \newblock {\em Beyond equilibrium thermodynamics}.
    \newblock John Wiley \& Sons, 2005.
    
    \bibitem{pavelka2020generalization}
    Michal Pavelka, V{\'a}clav Klika, and Miroslav Grmela.
    \newblock Generalization of the dynamical lack-of-fit reduction from generic to
      generic.
    \newblock {\em Journal of Statistical Physics}, 181(1):19--52, 2020.
    
    \bibitem{peletier2014variational}
    Mark~A Peletier.
    \newblock Variational modelling: Energies, gradient flows, and large
      deviations.
    \newblock {\em arXiv preprint arXiv:1402.1990}, 2014.
    
    \bibitem{seifert2005}
    Udo Seifert.
    \newblock Entropy production along a stochastic trajectory and an integral
      fluctuation theorem.
    \newblock {\em Physical Review Letters}, 95(4):040602, 2005.
    
    \bibitem{seifert2008stochastic}
    Udo Seifert.
    \newblock Stochastic thermodynamics: principles and perspectives.
    \newblock {\em The European Physical Journal B}, 64(3):423--431, 2008.
    
    \bibitem{seifert2012}
    Udo Seifert.
    \newblock Stochastic thermodynamics, fluctuation theorems and molecular
      machines.
    \newblock {\em Reports on Progress in Physics}, 75(12):126001, 2012.
    
    \bibitem{simo2006computational}
    Juan~C Simo and Thomas~JR Hughes.
    \newblock {\em Computational inelasticity}, volume~7.
    \newblock Springer Science \& Business Media, 2006.
    
    \bibitem{steele2001}
    J~Michael Steele.
    \newblock {\em Stochastic calculus and financial applications}, volume~1.
    \newblock Springer, 2001.
    
    \bibitem{still2012thermodynamics}
    Susanne Still, David~A Sivak, Anthony~J Bell, and Gavin~E Crooks.
    \newblock Thermodynamics of prediction.
    \newblock {\em Physical Review Letters}, 109(12):120604, 2012.
    
    \bibitem{torres2019combined}
    Alejandro Torres-S{\'a}nchez, Juan~M Vanegas, Prashant~K Purohit, and Marino
      Arroyo.
    \newblock Combined molecular/continuum modeling reveals the role of friction
      during fast unfolding of coiled-coil proteins.
    \newblock {\em Soft Matter}, 15(24):4961--4975, 2019.
    
    \bibitem{truesdell1984historical}
    Clifford Truesdell.
    \newblock Historical introit the origins of rational thermodynamics.
    \newblock In {\em Rational Thermodynamics}, pages 1--48. Springer, 1984.
    
    \bibitem{truskinovsky2005kinetics}
    Lev Truskinovsky and Anna Vainchtein.
    \newblock Kinetics of martensitic phase transitions: lattice model.
    \newblock {\em SIAM Journal on Applied Mathematics}, 66(2):533--553, 2005.
    
    \bibitem{turkington2013}
    Bruce Turkington.
    \newblock An optimization principle for deriving nonequilibrium statistical
      models of hamiltonian dynamics.
    \newblock {\em Journal of Statistical Physics}, 152(3):569--597, 2013.
    
    \bibitem{van2009unraveling}
    Joost van Mameren, Peter Gross, Geraldine Farge, Pleuni Hooijman, Mauro
      Modesti, Maria Falkenberg, Gijs~JL Wuite, and Erwin~JG Peterman.
    \newblock Unraveling the structure of dna during overstretching by using
      multicolor, single-molecule fluorescence imaging.
    \newblock {\em Proceedings of the National Academy of Sciences},
      106(43):18231--18236, 2009.
    
    \end{thebibliography}
\end{document}